# Unzipping Chemical Bonds of Non-Layered Bulk Structures to Form Ultrathin Nanocrystals

Liangxu Lin,[1,2*] Na Xu,[3] Chang Wu,[4] Juntong Huang,[5] Andrew Nattestad,[1] Xueling Zheng,[6] Gordon G. Wallace,[1] Shaowei Zhang[7,*] and Jun Chen[1,*]

[1]ARC Centre of Excellence for Electromaterials Science, Intelligent Polymer Research Institute, Australia Institute for Innovative Materials, Innovation Campus, University of Wollongong, Wollongong 2522, Australia
[2]Institute of Advanced Materials and Nanotechnology, The State Key Laboratory of Refractories and Metallurgy, Wuhan University of Science and Technology, Wuhan 430081, China
[3]College of Life Sciences and Health, Wuhan University of Science and Technology, Wuhan 430081, China
[4]Institute for Superconducting and Electronic Materials, Australian Institute for Innovative Materials, Innovation Campus, University of Wollongong, Wollongong 2522, Australia
[5]School of Materials Science and Engineering, Nanchang Hangkong University, Nanchang 330063, China
[6]College of Chemistry and Material Science, Fujian Normal University, Fuzhou 350007, China
[7]College of Engineering, Mathematics and Physical Sciences, University of Exeter, Exeter EX4 4FN, UK

*To whom correspondence should be addressed. E-mail: liangxu@uow.edu.au (L. L.), s.zhang@exeter.ac.uk (S. Z.), junc@uow.edu.au (J. C.).

Lead Contact: Liangxu Lin (liangxu@uow.edu.au).

## SUMMARY

The rich electronic and band structures of monolayered crystals offer versatile physical/chemical properties and subsequently a wide range of applications. Fabrication, particularly by the top-down "exfoliation" processes, rely on the presence of weak Van der Waals force between individual layers. Due to the strong chemical bonds between planes and atoms, un-zipping ultra-thin crystals (from one to several unit cells thick) from non-layered structures is more challenging. This work reports a technique which can be used to prepare such ultra-thin crystals from bulk non-layered structures ($a$-/$\beta$-MnO$_2$, ZnO, TiO$_2$, $a$-TiB$_2$). The physical and optical properties of these materials are characterized and contrasted against those from their bulk phases. The work presented here represents a tool kit for the preparation of novel 2D non-layered nanomaterials, providing significant contributions to this family of materials, paving the way for even more applications. Furthermore, we show the application of these novel NCs for bio-sensing and electrochemical oxygen reduction.

## INTRODUCTION

Both the fabrication and applications of two dimensional (2D) materials such as graphene and transition-metal chalcogenide (TMC) single layers have been intensively investigated in recent years.[1-13] Their properties (including luminescence, metallicity) are distinct from bulk phases, owing to the changed electronic and band structures.[1,2,6,8] Weak Van der Waals forces between layers mean exfoliation of 2D single layers and thin nanocrystals (NCs) from bulk layered materials are the most common approach for the production of such materials.[7-12] Meanwhile, other ultra-thin 2D non-layered crystals are predicted with unusual and potentially very useful properties. Some examples of such materials are 2D (011) crystal facet of rutile $TiO_2$ with a ~2.1 eV bandgap,[13] tunable metallic and insulating single layers of $MnO_2$ (only the $\delta$-$MnO_2$ is layered with different cell arrangement) in complex oxide heterostructure,[14] high temperature superconductivity and high electronic mobility of some single-layered metal oxide planes such as $Cu_2O$.[15,16] These novel physical and chemical properties are highly depended on the lattice arrangement, size confinement, defect and element doping at atomic level.[17] The plentiful range of crystal structures of non-layered materials offer versatile 2D interfaces with tunable properties.

The formation of such ultrathin layer is highly difficult (*e.g.* the precise control growth of single layer crystal facet, rarely reported) using traditional techniques such as ion sputtering and vacuum annealing,[13] pulsed laser deposition,[14,15] molecular beam epitaxy[16] and wet-chemical synthesis approach.[18] Liquid and chemical exfoliation processes, used with layered materials, are also unsuitable due to the strong chemical bond between crystal planes (Figure 1A,B).[7,9-11] A technique to unzip and disintegrate ultra-thin crystal planes from their bulk non-layered structures is of great interest toward the creation of novel 2D materials. To this end, we revisited the insertion/extraction process of several non-layered materials.[19,20] Alkali ions and atoms such as $Li^+$/Li and $Na^+$/Na can be inserted into the porous channels between/in unit cells (Figure 1C). Our previous studies showed that the reaction between the inserted K and ethanol/water disordered the surface and generated O vacancies of some metal oxide particles,[21] which can also break the covalent bond of C-C=C-C and B-N (with ionic bond characteristics) of layered materials.[8,22] Such reactions may be developed as a tool to unzip chemical bonds of non-layered structure, therefore to exfoliate and disintegrate bulk materials to produce ultra-thin crystals. The work here demonstrates how a "K-insertion and unzipping" technique can be developed for this purpose.

## RESULTS AND DISCUSSION

### Fabrication and Characterization

Figure 1 shows a schematic of our fabrication approach, using $\beta$-$MnO_2$ as an example. We start from a non-layered bulk $\beta$-$MnO_2$ (size > 20 μm, Figure 2A inset, scanning electron microscope/SEM image in Figure S1A) for the demonstration. Briefly, 1 g $\beta$-$MnO_2$ and ~0.5 g potassium metal (K) were transferred into a Pyrex tube which was then pumped to vacuum through a side connection. The Pyrex tube was heated at 190 - 200 °C for 4 h, during which the black particles changed appearance to dark brown. This heating allows the K insertion reaction to take place (Figure 1C, this reaction is further explored by X-ray powder diffraction (XRD) characterization below). Upon cooling to room temperature, $N_2$ flow was gently introduced through the side connection. Following this, 50 mL ethanol was poured into the tube followed by the addition of 50 mL deionized (DI) $H_2O$, and ultrasonicated for 2 h. A suspension was collected by centrifugation to remove the powder sediment. A strong cation exchange resin was used to remove $K^+$ from the raw suspension which was further centrifuged to remove any larger particles, giving a faint yellow NC suspension (Figure 1E-F, Figure 2A inset).

Unlike bulk $MnO_2$, the collected 2D-$MnO_2$ suspension was luminescent and has an absorption edge beyond 400 nm (Figure 2A, Tauc plot in Figure S2A). In addition to this, a shoulder is observed at around 280 nm. Following the sonication-centrifugation treatment (in EtOH/$H_2O$), the yield was calculated as around 4.1 wt%, based on the weight of the dried and purified suspension.

Nevertheless, the sediment can be collected and reprocessed (sonication-centrifugation process in pure water), leading to a total yield of ~ 36.8 wt% (Figure 2C - the concentrated suspension can be seen in Figure S3A). With this high yield the morphology of the final MnO$_2$ sediment was highly distinguishable from that of the bulk raw material (Figure S3B *versus* Figure S1A). The chemical structure of exfoliated NCs were characterized by XPS and Raman spectroscopy (Figure S3C-E).

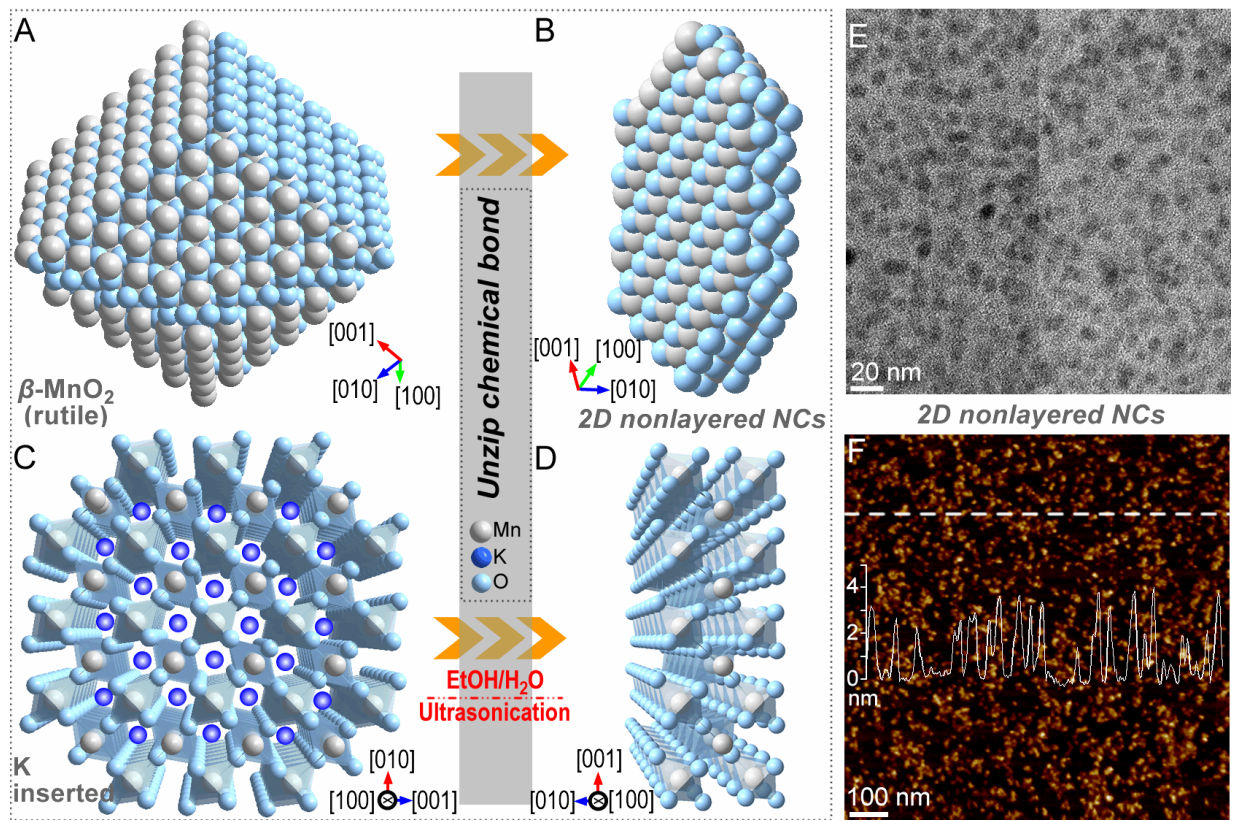

**Figure 1. Diagram shows the unzipping of $\beta$-MnO$_2$ to form 2D ultra-thin NCs**

(A-D) A and B are space filling mode, C and D are standard mode with polyhedron crystal structure. C shows the K inserted phase. The position of K atom is roughly arranged in the structure. Like Li$^+$ insertion of MnO$_2$,[20] these K atoms are also likely located at the channel position.

(E-F) (E) TEM and (F) atomic force microscopy (AFM) of the $\beta$-MnO$_2$ NCs (from the 1$^{st}$ fabrication). Inset in F is the line profile showing the thickness of NCs (high concentration, NCs were stacked).

X-Ray powder Diffraction (XRD) was employed to understand the detailed mechanism of how K inserts into and cleaves $\beta$-MnO$_2$. In this case, the amount of K for the reaction was reduced to ~0.3 g to decrease the magnitude of diffraction peaks from non-participant K. The dark brown K-inserted MnO$_2$ (K-MnO$_2$) powder was sealed by a thin polyimide film in the glove box to isolate from atmosphere (Figure 2B inset). A schematic structure of the $\beta$-MnO$_2$ with K insertions is shown in Figure 1C. The XRD trace of K-MnO$_2$ (Figure 2B, blue pattern) shows a near-total decrease in crystallinity, suggesting that K is able to fully diffused into and disorder the crystal structure of MnO$_2$ (ICCD card: 01-081-2261), however close examination of the XRD pattern of K-MnO$_2$ revealed many small diffraction peaks from the K inserted MnO$_2$ phases (Figure S3F). After the fabrication (the extraction of K), the crystal structure was gently recovered along with phase transformation to $a$-MnO$_2$ (ICCD card: 01-072-1982, red pattern). Our characterizations below suggested that the exfoliated NCs from the first time fabrication predominantly retain the $\beta$ phase. Therefore, this phase transformation was likely happened during the fabrication in the solvent rather the insertion reaction. There are no evident diffractions from $a$-MnO$_2$ identified in the XRD pattern of K-MnO$_2$ (Figure S3F). Furthermore, in the KOH solution (H$_2$O/EtOH solvent, further see Figure S3F) without the insertion reaction of K/K$^+$, the phase of $\beta$-MnO$_2$ was highly stable. A similar phase transformation was previously observed for alkaline ion insertion into MnO$_2$, which was a result of the thermodynamic phase selection.[25]

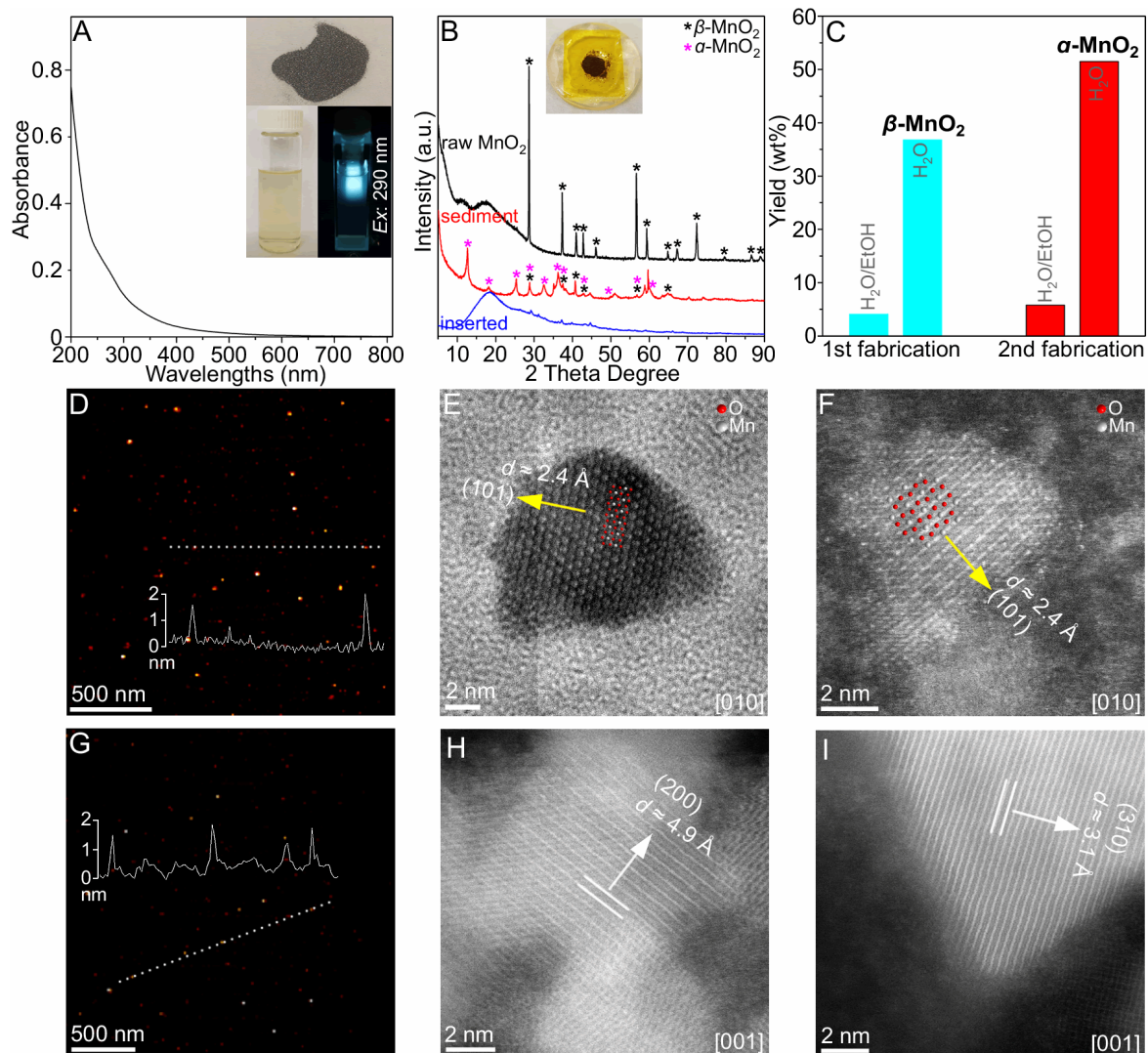

**Figure 2. Fabrication and characterization of ultrathin MnO$_2$ NCs**

(A) UV/Vis spectrum of 2D $\beta$-MnO$_2$ NCs suspension. Insets are bulk MnO$_2$ raw material, and pure $\beta$-MnO$_2$ NCs suspension with and without UV irradiation (excitation wavelength of 290 nm).

(B) XRD patterns of MnO$_2$ raw materials, sediment and K-MnO$_2$ with nearly same amount of the powder. Inset is the self-made XRD holder for the K-inserted sample, which has a XRD background below 20 °as indicated.

(C) The yield of MnO$_2$ NCs from 1$^{st}$ and 2$^{nd}$ fabrications, with and without further exfoliations in pure H$_2$O.

(D) AFM image of $\beta$-MnO$_2$ NCs from the 1$^{st}$ fabrication. The line profile shows the thickness of NCs.

(E-F) (E) HRTEM (bright field) and (F) aberration-corrected (AC) HRTEM (dark field) images of $\beta$-MnO$_2$ NC.

(G) AFM image of $a$-MnO$_2$ NCs from the 2$^{nd}$ fabrication. The line profile shows the thickness of NCs.

(H-I) AC-HRTEM (dark field) images show (200) and (310) lattices of $a$-MnO$_2$.

Atomic force microscopy (AFM) and transmission electron microscopy (TEM) characterizations confirmed the presence of ultra-thin (AFM thickness < 2 nm, similar to that of monolayered graphene oxide or monolayered transition metal chalcogenides[23,24] and thin NCs (Figure 1E,F, Figure 2D, size centered at ~15 nm, AFM size distribution in Figure S2B). HRTEM images of the NCs give clear lattice fringes of (101) $\beta$-MnO$_2$ (Figure 2E-F). With the assistance of fast Fourier transform patterns (Figure S4A-B), Mn and O atoms on the facet of $\beta$-MnO$_2$ can be seen to be well arranged in HRTEM images (Figure 2E-F). Additional HRTEM investigations (Figure S4C-H) suggested that the unzipping predominantly takes place parallel to the [001] zone axis of $\beta$-MnO$_2$ (*i.e.* (101) and (111) surfaces are exposed). Occasionally, some unzipped nanoribbons and ultrathin sheets with scrolled edge were identified (Figures S4I-L).

We repeated the K insertion and exfoliation process with the collected MnO$_2$ sediment powder, which was predominantly the *a*-MnO$_2$ phase (Figure 2B). Ultrathin NCs (Figure 2G) were also achieved with a yield of ~5.8 wt%. Similar to what was demonstrated in above, the final yield of NCs improved to ~51.5 wt% with repeated sonication and centrifugation, using pure water (Figure 2C). These NCs (Figure S5B-L, size centered at ~25 nm, AFM size distribution in Figure S2D) are slightly larger than the $\beta$-MnO$_2$ from the first fabrication (Figure S2B), and have similar Raman and XPS spectra (Figure S3C-E). On the other hand, the absorption differed slightly, with blue shifts in both the onset and the shoulder (to around 250 nm, Figure S5A, Tauc plot in Figure S2C). TEM suggested that NCs from this second fabrication process were predominantly *a*-MnO$_2$ with the majority of exposed surfaces perpendicular to the [001] zone axis (Figure S5B-L). AC-TEM images in Figure 2H-I clearly show the lattice spacings of (200) and (310) from *a*-MnO$_2$.

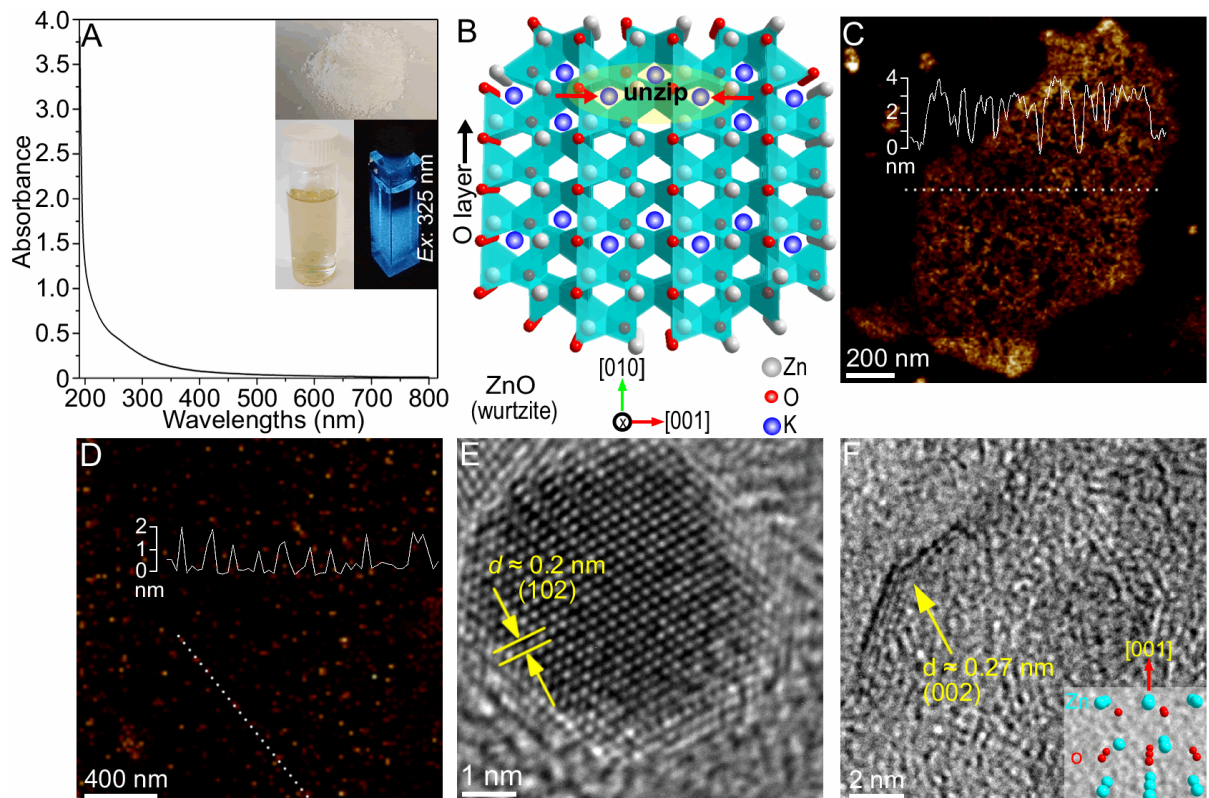

**Figure 3. Fabrication of ultrathin ZnO NCs**

(A) UV/Vis spectrum of 2D ZnO NCs suspension. Insets are photographs of the white ZnO raw material, pure ZnO NCs suspension with and without UV irradiation (excitation wavelength of 325 nm).

(B) Diagram shows the structure of wurtzite ZnO along [100] zone axis with some K atoms inserted.

(C-D) AFM images of the disintegrated (C) ZnO sheet and (D) NCs. The line profiles show the thickness of the sheet and NCs. (E-F) High-resolution TEM (HRTEM) images of ZnO NCs.

A similar process was conducted on bulk ZnO (size < 5 μm, Figure 3A inset, Figure S1B). Unlike the starting white ZnO powder, the collected ZnO suspension also had a faint yellow colour and was and luminescent (Figure 3A inset). The yield of this synthesis was calculated to be around 3.2 wt%, with XPS and Raman (Figure S6A-C) confirming the identity of the NCs to be ZnO, and to have a significant number of surface –OH groups. The featureless absorption spectrum in Figure 3A demonstrated high dispersibility of the ZnO product (Tauc plot in Figure S7A). XRD was also employed to examine the mechanism of K insertion into ZnO. A schematic of the process of K insertion into wurtzite ZnO is shown in Figure 3B. Again, the near total disappeared XRD diffraction peaks for K-ZnO (Figure S6D) shows that K able to fully diffuse into the ZnO crystal, in spite of the comparatively large size of ZnO crystals used here. Close examination reveals many small diffraction peaks from K inserted ZnO, which are ascribed to $K_2ZnO_2$ and $K_2Zn_6O_7$ (Figure S6E). K-ZnO was itself not particularly stable and changed appearance to gray with a partially restored crystal structure (Figure S6D) established during the fabrication and purification (repeated washing with DI water and ethanol to remove residual $K^+$). Nevertheless, AFM and TEM characterizations demonstrate that immediately following the reaction between K-ZnO and EtOH/$H_2O$ under ultrasonication, the material is layered, with Zn-O bonds being unzipped. Many large porous sheets with thicknesses < 4 nm were also found in the raw suspension ($K^+$ removed, but not well centrifuged), which were exfoliated from bulk ZnO and partially disintegrated (Figure 3C, Figure S6F). The majority materials in the purified and well centrifuged suspension were classified as thin (AFM thickness < 2 nm) and small (AFM centered at ~10-15 nm, Figure S8A-C, AFM size distribution in Figure S7B) and contained both Zn and O (Figure S8C-D). Lattice fringes of wurtzite ZnO were readily recognized in HRTEM images of these NCs (Figure 3E). Figure 3F shows a TEM image of one ZnO NC, laid perpendicular on the carbon film with 3-4 atomic layers of (002) lattice fringes, corresponding with one unit cell along the [001] zone axis. Further TEM investigation suggested that surface exposure perpendicular to the [010] zone axis (*i.e.* the exposure of (101) and (102) surfaces) was dominant (Figure S8E-I).

With similar fabrication protocols, ultra-thin 2D NCs were also produced from other metal oxides such as anatase $TiO_2$ (covalent compound, > 500 nm, Figure S1C) and even the metallic compound of *a*-$TiB_2$ (~ 10 μm, Figure S1D), with yields of ~ 3.6 wt% and 2.4 wt% respectively. The latter is a more complex structure as it contains a combination of covalent, ionic and metallic bonding. Aside from the surface –OH groups on both $TiO_2$ and $TiB_2$ NCs, XPS and Raman characterizations suggested that $TiB_2$ NCs were highly oxidized (Figure S9). Once again, these NCs were also luminescent, and have different absorptions compared to their bulk counterparts ($TiO_2$: faint yellow, $TiB_2$: light white, Figure 4A&D, Tauc plots in Figure S7C&E).

AFM imaging, shown in Figure 4, suggested these materials have roughly small sizes (average size around 10-15 nm of $TiO_2$, and < 10 nm of $TiB_2$, AFM size distribution in Figure S7D&F) and both ultrathin (AFM thickness < 2 nm). TEM further revealed $TiB_2$ NCs (~ 5-10 nm) to be much smaller than $TiO_2$ NCs (5-20 nm) (Figure S10, 11). The $TiO_2$ NCs showed regular atomic arrangements, with the majority of surfaces exposed being perpendicular to the [010] zone axis (Figure 4C, Figure S10). Some $TiO_2$ NCs included dislocations, giving slightly different stacking angles between the lattices of (200) and (004) (Figure S10K-L). By contrast, $TiB_2$ NCs was predominantly disordered (Figure S11), which was likely induced by the extensive oxidation during the fabrication process. Figure 4F shows one such disordered surface, with the (100) lattice spacings recognizable, but not clear. Although crystal surfaces perpendicular to the [010] zone axis were usually observed in NCs (Figure S11), we cannot conclude the main surface exposure of TiB$_2$ NCs. Like the above demonstrated K insertion process, the TiO$_2$ raw material was also fully disordered by K insertion, which then slightly recovered over time (Figure S12A,B). Due to the unzipping of ultrathin NCs, the exfoliated thin edges and highly cracked sheets were observed on some residual TiO$_2$ particles (Figure S13A-C). TiB$_2$ was the least amendable to insertion of K from the materials used here, which was determined through the existence of identifiable TiB$_2$ XRD peaks of the K-TiB$_2$ (Figure S12C,D). Nevertheless, residual TiB$_2$ (after five times fabrication) was also observed to have many thin crystal 'whiskers' on the surface (Figure S13E-I), suggesting infiltration but unsuccessful disintegration and is distinct from images of unprocessed TiB$_2$ (Figure S13D). The yield was further improved with subsequent re-processing of the residual powder (Figure S14A). With more fabrications, the yield of ZnO and TiO$_2$ was slightly increased, whilst the yield of TiB$_2$ was relatively stable, which may because of the relative more stable phase structure of the TiB$_2$ over that of TiO$_2$ and ZnO.

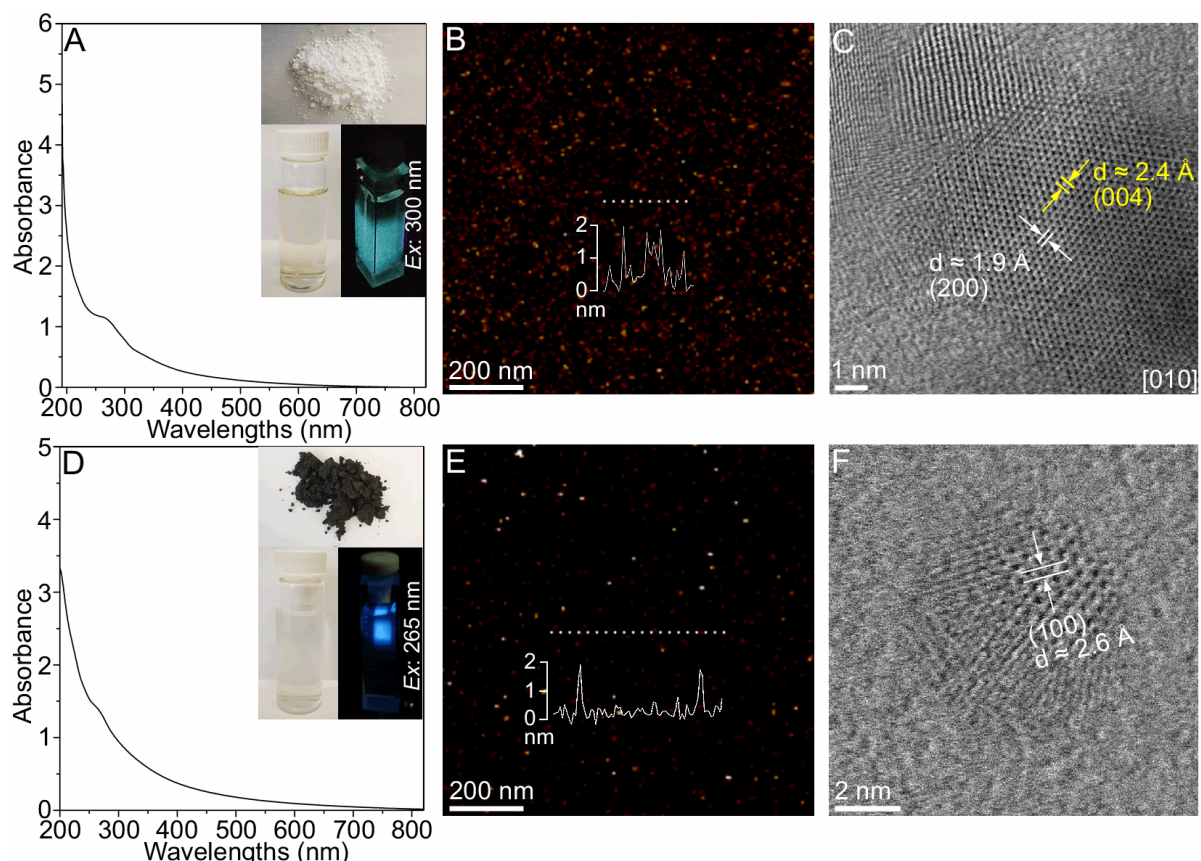

**Figure 4. Fabrication of ultrathin TiO$_2$ and TiB$_2$ NCs**

(A) UV/Vis spectrum of 2D TiO$_2$ NCs suspension. Insets are bulk TiO$_2$ raw material, and pure TiO$_2$ NCs suspension with and without UV irradiation (excitation wavelength of 300 nm).

(B) AFM image of TiO$_2$ NCs. The line profile shows the thickness of NCs.

(C) HRTEM (bright field) image of one TiO$_2$ NC.

(D) UV/Vis spectrum of 2D TiB$_2$ NCs suspension. Insets are bulk TiB$_2$ raw material, and pure TiB$_2$ NCs suspension with and without UV irradiation (excitation wavelength of 265 nm).

(E) AFM image of TiB$_2$ NCs. The line profile shows the thickness of NCs.

(F) HRTEM (bright field) image of one TiB$_2$ NC.

All these fabricated NCs are highly stable in aqueous phase. As for the fabrication technique, it remains highly challenging to directly grow such crystallized ultrathin 2D materials using bottom-up techniques such as molecular beam epitaxy and wet-chemical synthesis.[13-18] Meanwhile, top down strategies such as liquid and chemical exfoliations have been proven successful for some ultrathin 2D materials, however this relies on the layered materials having weak van der Waals forces between layers.[7,26,27] Other exfoliation techniques have also been demonstrated with different chemical intercalants (*n*-butyllithium, multi-component approaches), however these unfortunately still can only exfoliate 2D layers from layered structures, without strong chemical bonds between these layers.[9,11] Importantly our work here suggest that the developed "K-insertion and unzipping" technique can be adapted to unzip many bulk, non-layered, materials with chemical bonds ranging from covalent to ionic and metallic bonds. This allows us to form 2D ultrathin structures, providing an avenue for various novel 2D non-layered materials, interfaces, and associated applications. The technique has the potential to be used to unzip large 2D ultrathin sheets (*e.g.* Figure S4K and 6F). Specific size distribution of 2D materials could be reached by gradient centrifugation and different dialysis treatments. Nevertheless, we tried to reduce the ultrasonication power (reduced to 80%) for the identical fabrication of 2D $\beta$-MnO$_2$. The size in this case was roughly improved as around 20-100 nm, while the NCs were still really thin (Figure S14B,C). We have further conducted the unzipping reaction of the MnO$_2$ with Na metal (0.3 g Na and 1 g $\beta$-MnO$_2$). The Na can also be inserted, although it was relatively difficult to wet the $\beta$-MnO$_2$ powder compared to the K. The Na insertion also highly disordered the crystal structure of $\beta$-MnO$_2$, which was transformed to the *a*-phase after the fabrication in the solvent (Figure S14D). Our characterizations suggested that 2D ultrathin crystal edge (nanoribbon: width of over 10 nm, length of up to hundreds nanometers) as well as some small NCs (mainly *a*-MnO$_2$) have been exfoliated (Figure S15). Occasionally, some 2D sheets with larger size were also found in the suspension (Figure S15E,F). Compared to K, the Na atom has a smaller radius, which is also relatively gentler in reacting with EtOH/H$_2$O solvent. It means that the mitigation of the unzipping process could also be potentially used to further control the lateral size and morphology of 2D non-layered materials. Nevertheless, we not suggest the using of Li for the unzipping reaction due to the superior diffusion ability of the Li (*e.g.* diffuse into the Pyrex tube). Below, the optical properties of these ultrathin 2D NC structures are reported, along with evaluation of these materials for use in ORR and biosensing applications.

**Optical Properties of 2D non-Layered NCs**

As all the reported NCs (prepared with K intercalant) were strongly luminescent, quantum yields (PLQY) were able to be readily determined, using a fluorescence spectrometer with an integration sphere. PLQY values of 8.37%, 5.42%, 7.16%, 9.21% and 2.15% were observed for $\beta$-MnO$_2$, *a*-MnO$_2$, ZnO, TiO$_2$ and TiB$_2$ NCs, respectively. Differences in the surface facet exposure and crystal phase of MnO$_2$ lead to significant changes on optical properties. As shown in Figure 5A, the maximum photoluminescence (PL) emissions of $\beta$-MnO$_2$ and *a*-MnO$_2$ NCs were found at 400 and 430 nm with peak PL excitation (PLE) wavelengths of 290 and 325 nm, respectively. $\beta$-MnO$_2$ NCs have a PLE maxima at ~290 nm, with a weak shoulder at ~340 nm, while *a*-MnO$_2$ NCs showed PLE peaks at 260 and 325 nm. Quantum confinement effects were ruled out as the origin of these differences, as TEM characterization showed no evidence of smaller crystal sizes for the *a*-MnO$_2$ sample as compared with $\beta$-MnO$_2$ NCs. Furthermore, the PL kinetics (Time Resolved PL, TRPL) of these NCs were distinct from one another. *a*-MnO$_2$ was fitted with three components (0.7, 3.6 and 11.4 ns), while only two components (2.2 and 6.8 ns) were seen for of $\beta$-MnO$_2$ (Figure S14E-F). Similarly, optical properties of the other fabricated NCs were characterized in detail. ZnO NCs have an optical bandgap of ~4.83 eV and an in-band gap of ~ 3.82 eV, which is significantly larger than the direct gap of previously reported ZnO quantum dots (3.7-3.9 eV, size around 3-4 nm, Bohr radius ~ 1.8 nm) (Figure 5B).[28,29] The ZnO emission peak was at ~440 nm when excited at 325 nm, life times of 2.9 and 9.6 ns (70 : 29 % ratio, Figure S14G), which are significantly longer than other ZnO nanomaterials where hundreds of picoseconds have been reported.[30,31] As for the TiO$_2$ and TiB$_2$ NCs, they have emission peaks at ~415 nm (excitation of 300 nm) and ~365 nm (excitation of 260 nm) respectively (Figure 5C-D), and luminescent decay time at nanosecond level (Figure S14H-I). As mentioned, the band structure of $a$-TiB$_2$ NCs was more complex because of the surface oxidization, giving likely new emission centers (*e.g.* B-O, Ti-B and 1,3-B centers),[8,32-33] and resulting three gap levels at around 5.49, 4.85 and 4.15 eV (versus 4.88 and 4.04 eV of TiO$_2$). Precise assignments of the effects on these optical properties are challenging and the subject of ongoing work. Nevertheless, the above data already demonstrates that novel optical properties resulted from the creation of ultra-thin 2D structures. These are decided by not only the surface exposure, crystal phase, but also the other surface arrangement such as doping and oxidization.

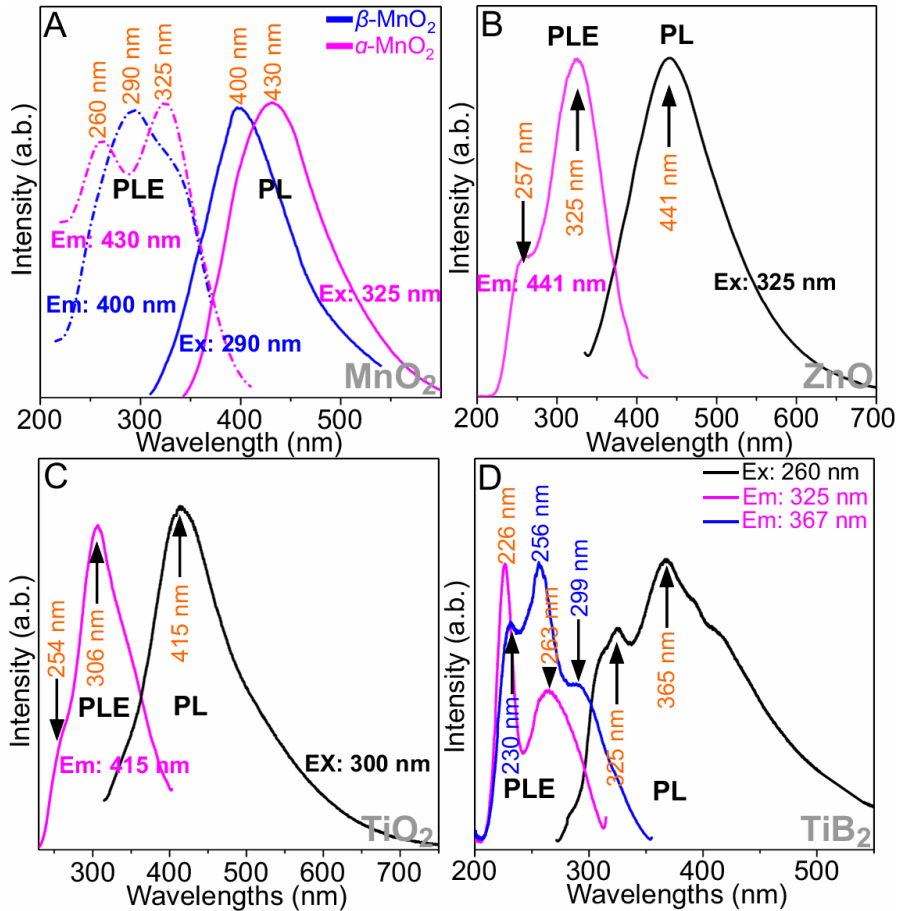

**Figure 5. PL and PLE spectra of the fabricated 2D NCs**
(A) $a$- and $\beta$-MnO$_2$ NCs. The emission (for PLE) and excitation wavelengths (for PL) are recorded in the figure. (B-D) (B) ZnO, (C) TiO$_2$ and (D) TiB$_2$ NCs.

## Applications of 2D non-Layered NCs

These 2D non-layered structures may have already find promising applications in different fields. An immediate application of the NCs is the bio-sensing by using the various luminescent emissions with nanosecond decay kinetics. By way of example, we demonstrate TiB$_2$ NCs as biosensors for four different tumor cells (HUVEC, HeLa, NIH-3T3 and MC3T3-E1). These cells were cultured and plated at 1 ×10$^5$ cell per well across 24-well plates for 24 h. TiB$_2$ NCs with different concentrations (50, 100, 150 and 200 µg mL$^{-1}$, see Experimental Procedures for confirmation of NC concentrations) were added to each well before being incubated at 37 $^o$C for 24 h. Cell counting kit-8 (CCK8) was used to determine the number of viable cells in the cytotoxicity assays. Figure 6A shows the cytotoxicity result of TiB$_2$ NCs, which suggested that NCs were non-toxic to all these four cells at the concentrations of 100 µg mL$^{-1}$. We then recorded the confocal microscopy images of the HUVEC cell once it was incubated in 50 µg mL$^{-1}$ NCs for 24 h and washed by PBS solution. The control image (no luminescence excited) of the cell was not well defined (Figure 6B), however emission from photoexcited NCs showed clear boundaries between cells, as well as identifying nuclei and cytoplasm (Figure 6C). Differences between cell nuclei and cytoplasm were more clearly shown in the merged image, with different light excitations, when the cell was also stained by DAPI (4',6-diamidino-2-phenylindole, Figure 6D). Images of other cells gave similar results (Figure S16), and suggest that TiB$_2$ NCs can be potentially used in high contrast bio-imaging applications. Upon suitable surface modifications and size selections, interesting bio-interactions with similar 2D interfaces (*e.g.* MnO$_2$) are also possible.[35]

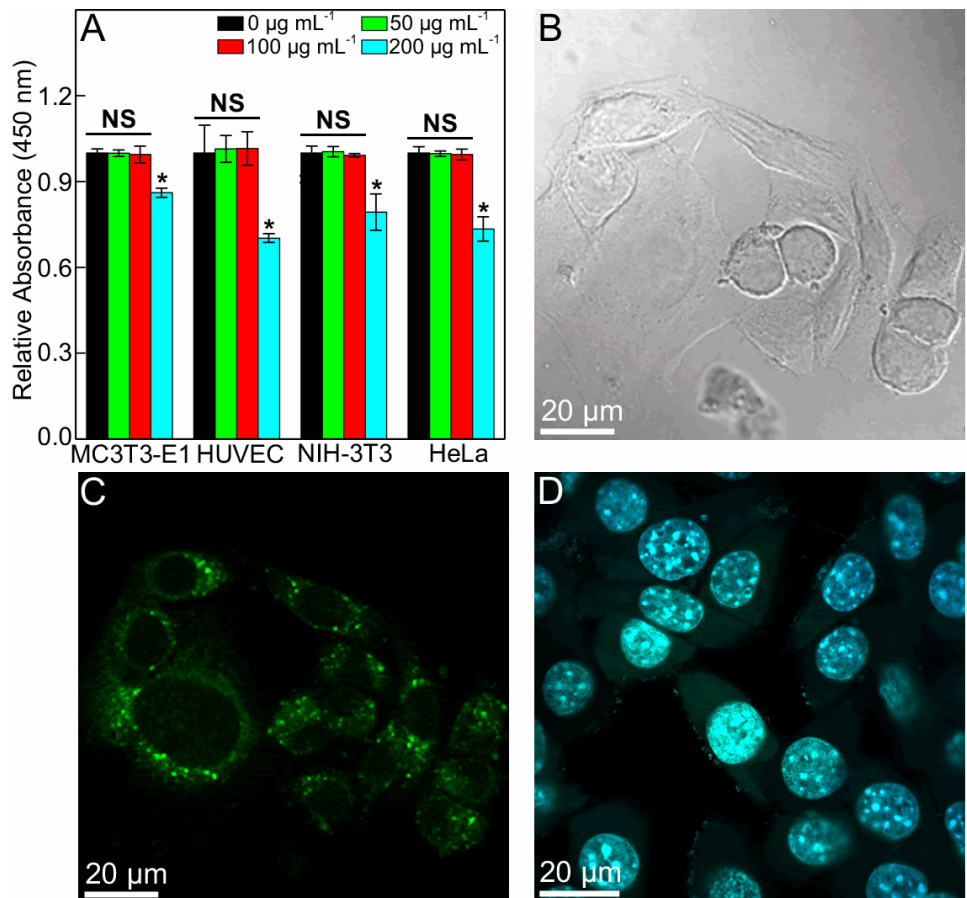

**Figure 6. Bio-sensing applications of TiB$_2$ NCs.**
(A) Cell viability assay of different cells incubated (24 h) with TiB$_2$ NCs at different concentrations (NS: no significant difference).
(B-C) Confocal microscopy images of HUVEC cells stained with TiB$_2$ NCs (B) without and with (C) single light excitation.

(D) Confocal microscopy image of HUVEC cells stained with both DAPI and $TiB_2$ NCs (merged from two images with different light excitations).

These NCs may also be employed in electrocatalysis, as various non-layered materials have proven performance in a range of catalytic reactions, such as oxygen (ORR) and $CO_2$ reduction ($CO_2$RR).[36-40] Nevertheless, electrocatalysis is known to be limited by the inefficient charge transfer.[39-43] In this regard, the formation of ultrathin 2D structures is promising, as this provides much short paths for charges to reach surfaces, along with the rich surface chemistry for which can be tuned for electrocatalysis. We therefore screened $a$-$MnO_2$ NCs for ORR in a 1 M KOH electrolyte, saturated with $O_2$. Results of this preliminary measurement are shown in Figure S17. To overcome stacking and agglomeration of 2D NCs, which detrimentally affected the linear sweep voltammetry (LSV), we dipped carbon paper (area of ~1 $cm^2$) in the diluted suspension (~500 µg $mL^{-1}$), which was then dried. The background current observed in the LSV of this carbon paper was negligible (Figure S17A), while bulk $MnO_2$ at similar loadings, showed some response. On the other hand, $a$-$MnO_2$ NCs exhibited significantly improved current limit of around 6.3 mA $cm^{-2}$, at around 0.45 V $vs$ RHE. Onset and half-wave potentials were observed at around 0.8 and 0.53 V $vs$. RHE, respectively (Figure S17A). This catalysis is already efficient compared with other reports of $MnO_2$ in a static arrangement,[39,43,44] however, in order to record the electron transfer number, 10 µL NCs were also immobilized on a rotating ring-disc electrode (RRDE, 1600 R.P.M., collection efficiency of 25.6%). A voltage of 1.3 V was applied on the ring, to reduce $H_2O_2$ generated during cyclic voltammetry (CV). In the potential range of our test, the electron transfer number and production yield of $H_2O_2$ were determined to be around 2.94 and 50%, respectively (Figure S17B,C). The catalytic efficiency estimated with the RRDE may be an under-estimate, due to stacking of NCs, as we previously discussed in detail for another 2D material system.[23] Nevertheless, the above measurements suggest that these 2D NCs are promising candidates for selective and efficient electrocatalysis if their surface can be further regulated, controlling the chemical affinity to $O_2$ and electronic distribution $via$ chemical reduction and chemical doping on the 2D surface.

Based on the above applications, we foresee a great future for these and other 2D non-layered materials. This synthesis approach is a toolbox to enable various implementations ($e.g.$ edge and surface polarized electronic structure, catalysis, energy storage and bio-science). The successful fabrication of these novel 2D interfaces offers great opportunities to such research and applications.

**Conclusion**

We have demonstrated a novel "K-insertion and unzipping" technique to fabricate ultrathin 2D NCs with thicknesses of up to around 2 nm measured by AFM, corresponding to a single crystal unit cell, as well as some nanosheets, from bulk non-layered structures of wurtzite ZnO, $\beta$-$MnO_2$, anatase $TiO_2$ and $a$-$TiB_2$. This fabrication process involved the simple insertion of K into different non-layered structures $via$ a heating reaction under vacuum condition, followed by ultrasonication in $EtOH/H_2O$ at room temperature. The K insertion lead to fully disordered structures, however the crystal structures were recovered upon K removal, following successfully unzipping of the chemical bond. Through repeated processing, significant yields were attained. The above fabrication processes, leading to 2D NCs and nanosheets have been verified by various characterization techniques. Our investigations also suggested the phase transformation of $MnO_2$ materials during the fabrication, giving NCs with both $\beta$ and $a$ phases. Owing to the rich surface chemistry, the successfully fabricated novel 2D non-layered NCs yielded different electronic structures, strengthening the family of 2D nanomaterials, giving potentially various useful applications. As proof-of-concept studies, we demonstrated that the $TiB_2$ and $a$-$MnO_2$ NCs can be well used in bio-sensing and ORR electrocatalysis applications. Future works may include the use of various 2D non-layered interfaces with particular controlled and tailored surface structures for more promising energy conversion and bio-science applications.

## EXPERIMENTAL PROCEDURES

### Resource Availability

*Lead Contact*
Further information and requests for resources and reagents should be directed to and will be fulfilled by the Lead Contact, Liangxu Lin (liangxu@uow.edu.au).

*Materials Availability*
This study did not generate new unique reagents.

*Data and Code Availability*
This study did not generate/analyze any datasets/code.

### Materials and Fabrication

ZnO NCs: In the glove-box, 1 g ZnO (< 5 μm, Sigma Aldrich: product No. 205532) and ~0.5 g K (Sigma Aldrich: product No. 244864) were transferred into a Pyrex tube which was then mounted in a fume cupboard and pumped to vacuum through the side connection. The Pyrex tube was first heated to 140 $^o$C and gently shaken to allow K to fully melt and mixed with ZnO. The tube was further heated to190 - 200 $^o$C and kept for 4 h. After the reaction, the Pyrex tube was cooled and $N_2$ was gently introduce into the tube through the side connection. Subsequently, 50 mL ethanol and then 50 mL deionized (DI) $H_2O$ were poured into the Pyrex tube, which was then ultrasonicated for 2 h (Bandelin DL102H, 480 W, 35 kHZ). A suspension was collected by simple centrifugation to remove powder sediment. The sediment was ultrosonicated again in 100 mL EtOH/$H_2O$ (1: 1 volume ratio), and the suspension was collected by centrifugation. The suspension was repeatedly washed by a strongly acidic cation exchanger to remove any residual $K^+$. This suspension was further centrifuged to remove any large particles, resulting finally in a faint yellow NC suspension. For the repeated processing, the collected ZnO sediment was washed by DI water and ethanol, dried in the oven, and then cooled down to room temperature in a desiccator. The collected sediment was mixed with ~0.5 g K in the Pyrex tube again for the repeated reaction as that described above. The yield of the NCs was calculated by the weight of the fully dried suspension (~80% dried and weighed, and ~20% remained for other characterizations).

$MnO_2$, $TiO_2$, $TiB_2$ NCs: Materials were purchased from the Sigma-Aldrich ($TiB_2$: product No. 336289; $MnO_2$: product No. 243442; $TiO_2$: product No. 248576). Fabrication of these NCs were followed the similar procedure with that of ZnO NCs. Wetting of $TiB_2$ by molten K was difficult and the reaction was performed at over 200 $^o$C for > 6 h with occasional shaking of the tube. The repeated processing of the $TiO_2$ and $TiB_2$ NCs was the same as for ZnO, while for $MnO_2$, the residual material was dispersed in 100 mL DI water and was treated by ultrasound for 30 mins to further exfoliate NCs. Collection of the NCs was the same as that for the initial processing. The exfoliation process was repeated 8 times, and residual $MnO_2$ was further mixed with K with the same stoichiometric ratio to that of the first-time fabrication. Separation and purification of the NCs were also the same to that of ZnO. The yield of the NCs was also calculated by the weight of the fully dried suspension.

## Characterization

SEM was performed using a JSM-7500F field emission scanning electron microscope. XRD was performed using a PANalytical Empyrean XRD (Cu Kα, wavelength ~ 0.154 nm, with a scanning step of $2^o$ min$^{-1}$). PL and PLE spectra were recorded using a Hitachi F-4500 Fluorescence Spectrophotometer at 20 $^o$C. TRPL spectra and PLQY were recorded using a Horiba Fluorolog-3 fluorescence spectrometer. For TRPL, 340 nm excitation was provided from a NanoLED and corresponding maximum emissions at different wavelengths were recorded. Quantum yield was recorded using a Quanta-phi integration sphere attachment and performed at room temperature, with dilute suspensions of NCs. Atomic force microscopy (AFM) was performed using a VEECO Dimension 3100 system in tapping mode with a scan rate of 1 Hz (with samples placed on mica substrates). TEM and AC-TEM images were obtained using a JEOL JEM-2010 and JEOL ARM 200F at 200 kV, respectively. Absorption spectra were recorded using a Shimadzu-3600 UV-VIS-NIR spectrophotometer. Since direct weighing of NCs was difficult, we calculate the concentration of the suspension from the weight of the other part dried NCs. We also recorded the UV/Vis absorption intensity of the suspension, and the absorption co-efficiency (to the mass) was used to verify and regulate the concentration of NCs for bio-imaging experiment and catalysis test. XPS was performed on a Thermo Fisher Nexsa X-Ray Photoelectron Spectrometer (XPS) System. The binding energy of XPS was calibrated as 284.6 eV of the C1s. Raman spectra were recorded by the Horiba LabRAM HR Evolution with excitation wavelength of 633 nm.

## Bio-imaging experiments

Cells were cultured and maintained in 1 mL DMEM (Dulbecco's Modified Eagle Medium) media, with 10% fetal bovine serum, 100 IU/mL penicillin and 100 μg/mL streptomycin), and plated at $1 \times 10^5$ cell per well (CITOGLAS®10212450C cover glass inside) on 24-well plates for 24 h. TiB$_2$ NCs with different concentrations (50, 100, 150 and 200 μg mL$^{-1}$) were added to each well, and the cell was incubated at 37 $^o$C for 24 h. The number of viable cells in the cytotoxicity assays was recorded by a micro-plate reader at 450 nm with CCK-8. For the bio-imaging, the cells on cover glasses were washed with 1 mL $1 \times$ PBS, then were fixed using 4% paraformaldehyde at room temperature for 5 min. Cellular images were taken using the Olympus FV3000 confocal microscope. As an alternative method, the fixed cells were also washed by DAPI (Beyotim C1006, China) to stain cell nucleus. Cellular images were performed under different light excitations, and the obtained images were merged to distinguish the cell cytoplasm and nucleus mote clearly.

## Electrochemical measurements

Electrochemical measurements were performed on a Biologic VSP-300 electrochemical workstation with a three electrode set up. Hg/HgO was used as a reference while a carbon rod was employed as a counter electrode and the working electrode was on carbon paper (~ 1 cm$^2$) with carbon electrode clamp. For RRDE experiments with the PINE WaveNow Potentiostat/Galvanostat System, a glassy carbon counter electrode, Pine E6R1 RRDE (collection efficiency of 25.6%) was used. LSVs were recorded at a scan rate of 5 mV s$^{-1}$ in a 1M KOH (pH = 14) electrolyte with saturated O$_2$ and $iR$ corrections. Potentials were calibrated to a reversible hydrogen electrode (RHE) with the Nernst equation: E(RHE) = E(Hg/HgO) + 0.059 × (pH) + 0.098. To record the current at the ring, 10 μL NCs (~500 μg mL$^{-1}$) NCs were immobilized onto the glassy carbon of the RRDE. CV scan was applied at 1600 RPM with 1.3 V applied on the ring, and the current density at both the glassy carbon and ring were collected. Electron transfer number and production yield of $H_2O_2$ were subsequently calculated. Considering the influence of the stacking of 2D NCs, we also immersed a carbon paper into the NCs suspension and dried for LSV measurement. Raw $MnO_2$ materials were also loaded on the carbon paper with similar concentration of the suspension.

## SUPPLEMENTAL INFORMATION

Supplemental Information can be found online at https://doi.org/xxxxxx.

## ACKNOWLEDGEMENT

We thank Prof. Huihua Min at Nanjing Forestry University and Prof. G. Liang at Hubei University of Arts and Science for the assistance on TEM and basic PL characterizations. Financial supports from UOW VC Fellowship and ARC Centre of Excellence Scheme (CE140100012) are gratefully acknowledged. The authors would like to thank the Australian National Fabrication Facility – Materials node for facility access.

## AUTHOTR CONTRIBUTION

L.L. convinced the research and conducted the main experiments. L.L. also supervised the project, while J.C, S.Z. and G.G.W. provided oversight to the experimental work and drafted the manuscript with L.L. N.X conducted the bio-experiments. C.W, J.H., A.N. and X. Z. assisted on the XRD and catalysis measurements, the early stage fabrications, the optical measurements and AFM characterizations, respectively.

## DECLARATION OF INTERESTS

The authors declare no competing interests.

# Unzipping Chemical Bonds of Non-Layered Bulk Structures to Form Ultrathin Nanocrystals

Liangxu Lin,[1,2*] Na Xu,[3] Chan Wu,[4] Juntong Huang,[5] Andrew Nattestad,[1] Xueling Zheng,[6] Gordon G. Wallace,[1] Shaowei Zhang[7,*] and Jun Chen[1,*]

[1]ARC Centre of Excellence for Electromaterials Science, Intelligent Polymer Research Institute, Australia Institute for Innovative Materials, Innovation Campus, University of Wollongong, Wollongong 2522, Australia

[2]Institute of Advanced Materials and Nanotechnology, The State Key Laboratory of Refractories and Metallurgy, Wuhan University of Science and Technology, Wuhan 430081, China

[3]College of Life Sciences and Health, Wuhan University of Science and Technology, Wuhan 430081, China

[4]Institute for Superconducting and Electronic Materials, Australian Institute for Innovative Materials, Innovation Campus, University of Wollongong, Wollongong 2522, Australia

[5]School of Materials Science and Engineering, Nanchang Hangkong University, Nanchang 330063, China

[6]College of Chemistry and Material Science, Fujian Normal University, Fuzhou 350007, China

[7]College of Engineering, Mathematics and Physical Sciences, University of Exeter, Exeter EX4 4FN, UK

*To whom correspondence should be addressed. E-mail: liangxu@uow.edu.au (L. L.), s.zhang@exeter.ac.uk (S. Z.), junc@uow.edu.au (J. C.)

**Supplemental data items**

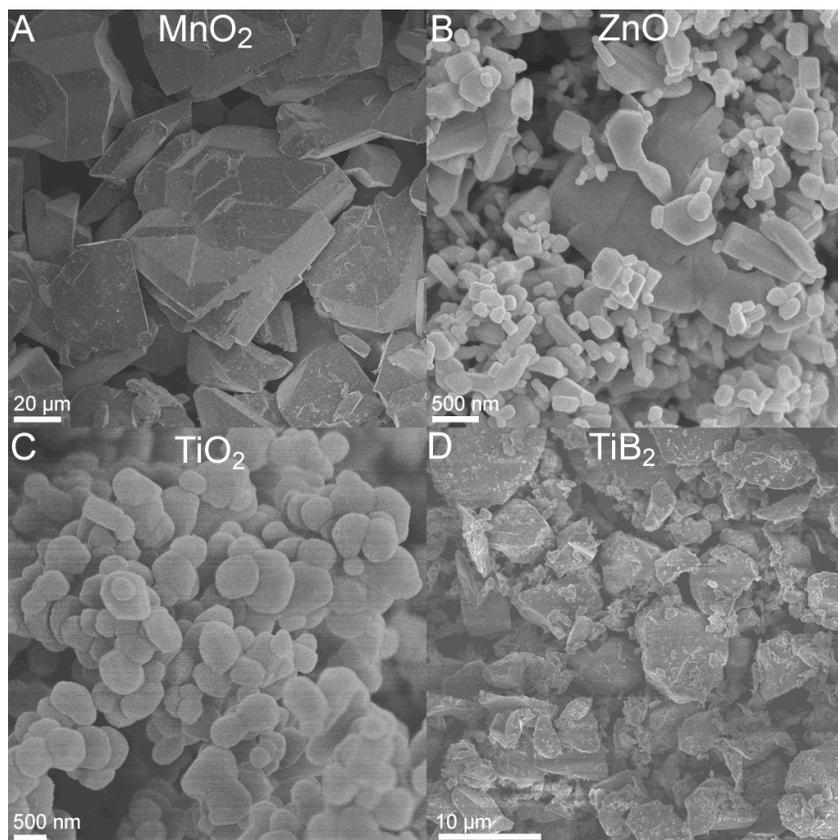

**Figure S1. SEM images of raw materials involved in this work.**
(A) $MnO_2$.
(B) ZnO.
(C) $TiO_2$.
(D) $TiB_2$.

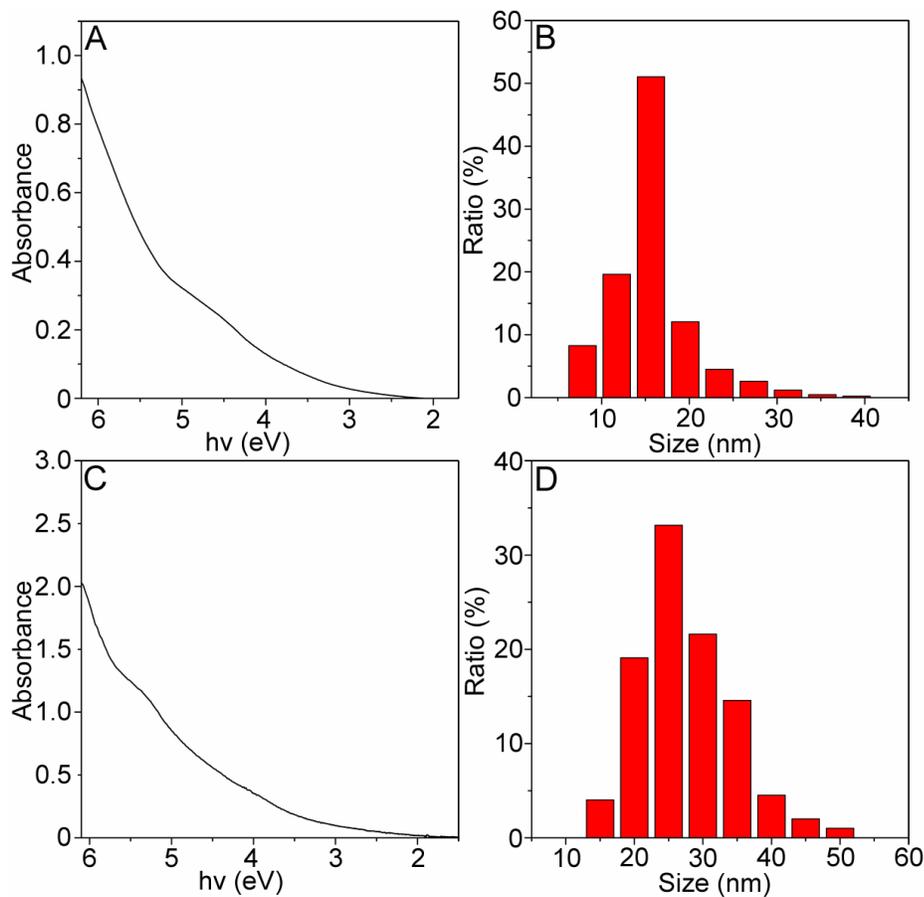

**Figure S2. UV/Vis spectrum (Tauc plot) and size distribution of MnO$_2$ NCs**

(A) Tauc plot of $\beta$-MnO$_2$ NCs.

(B) Size distribution of $\beta$-MnO$_2$ NCs (from AFM image, the size from AFM image may be larger than the real size of NCs.).

(C) Tauc plot of $a$-MnO$_2$ NCs.

(D) Size distribution of $a$-MnO$_2$ NCs (from AFM image, the size from AFM image may be larger than the real size of NCs.).

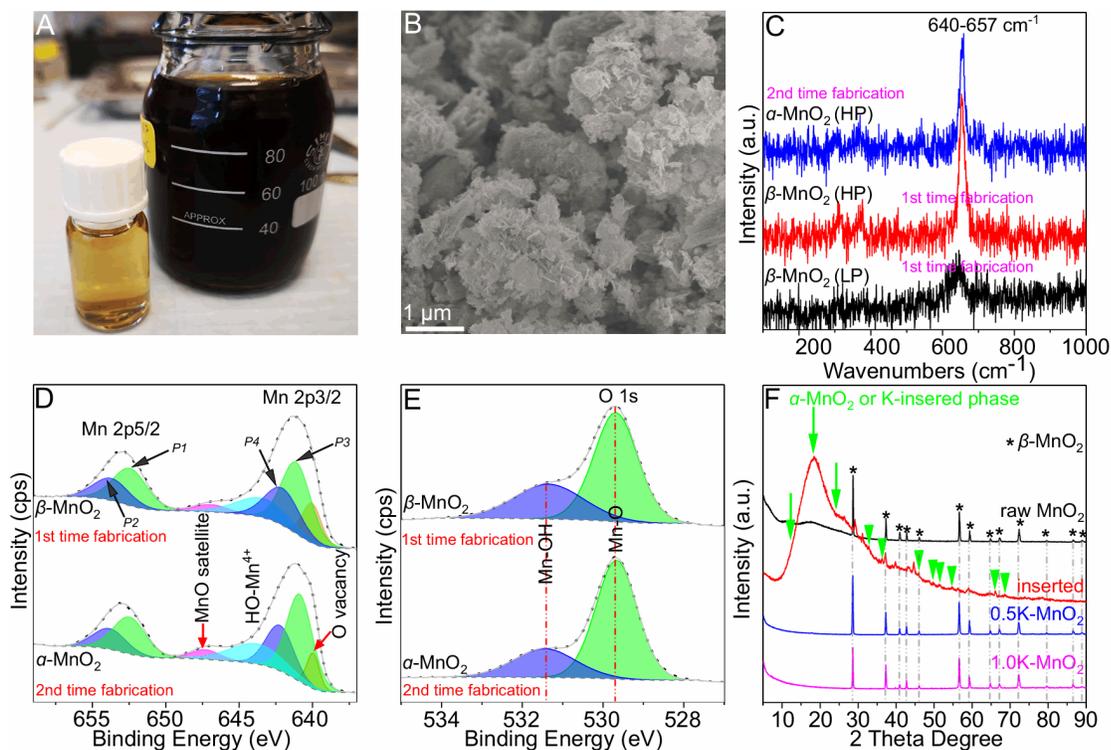

**Figure S3. Fabrication and characterization of MnO₂ NCs**

(A) The concentrated and diluted $\beta$-MnO₂ NCs suspension.

(B) SEM image of the residual MnO₂ material after one time fabrication.

(C) Raman spectra of $\beta$-MnO₂ (the 1ˢᵗ time fabrication) and $a$-MnO₂ NCs (the 2ⁿᵈ time fabrication).

(D-E) Mn2p (C) and O1s XPS (D) of MnO₂ NCs from the 1ˢᵗ ($\beta$ phase NCs) and 2ⁿᵈ ($a$ phase NCs) fabrications.

(F) XRD patterns of raw MnO₂, K-MnO₂, 0.5K-MnO₂, 1.0K-MnO₂. XRD intensity was changed for comparison.

**Additional information of Figure S3**: Raman response under low powder (LP) laser irradiation was really weak (C). Therefore, the beam irradiation with high power (HP) was applied. The peaks at~640-657 $cm^{-1}$ are the stretching modes of MnO₆ octahedra.[S1] The XPS peak at 643.9 and 640.1 eV (D) are the Mn states binding with –OH and with O vacancies, respectively.[S2] $\beta$- and $a$-MnO₂ NCs have satellite peaks of MnO at ~647.1 and 647.4 eV respectively.[S3] The split-orbit components of $\beta$- and $a$-MnO₂ were 11.66 and 11.75 eV, respectively. The Mn2p of $\beta$-MnO₂ can be fitted as $P_1$, $P_2$, $P_3$ and $P_4$ at 652.6, 653.9, 641.2 and 642.3 eV, respectively. P1-P4 of $a$-MnO₂ are 652.6, 654.0, 640.9 and 642.3 eV, respectively. $P_1/P_3$ and $P_2/P_4$ were ascribed to the loose surface structure and the body MnO₆ octahedra, respectively. These peaks of NCs were all down shifted to the lower binding energies comparing with nanofibers with (110) exposures,[S4] suggesting that the ratio of loose surface structure in our NCs was significant higher, corresponding well with the thin structure. The difference between Mn2p of $\beta$- and $a$-MnO₂ NCs is due to the different crystal structures and surface exposures. In O1s XPS (D), the Mn-O and O species (from the surface -OH) of both $\beta$- and $a$-MnO₂ NCs were found as 529.4 and 531.4 eV, respectively.[S4] In XRD patterns (F), the really weak labeled peaks in K-MnO₂ might be $a$-MnO₂, which can also be assigned to the K inserted phase (*e.g.* ICCD card: 00-042-1317, 00-016-0205 and 00-012-0706). As the control samples, MnO₂ was ground and dispersed in 100 mL H₂O/EtOH solvent (1 : 1 volume ratio). KOH with equivalent molar quantity of 0.5 and 0.25 g K was added in to the suspension, which was then sonicated for 2 h and dispersed for 3 days, giving samples of 0.5K-MnO₂ and 1.0K-MnO₂, respectively. XRD of these samples suggested that the K⁺ did not change the crystal phase.

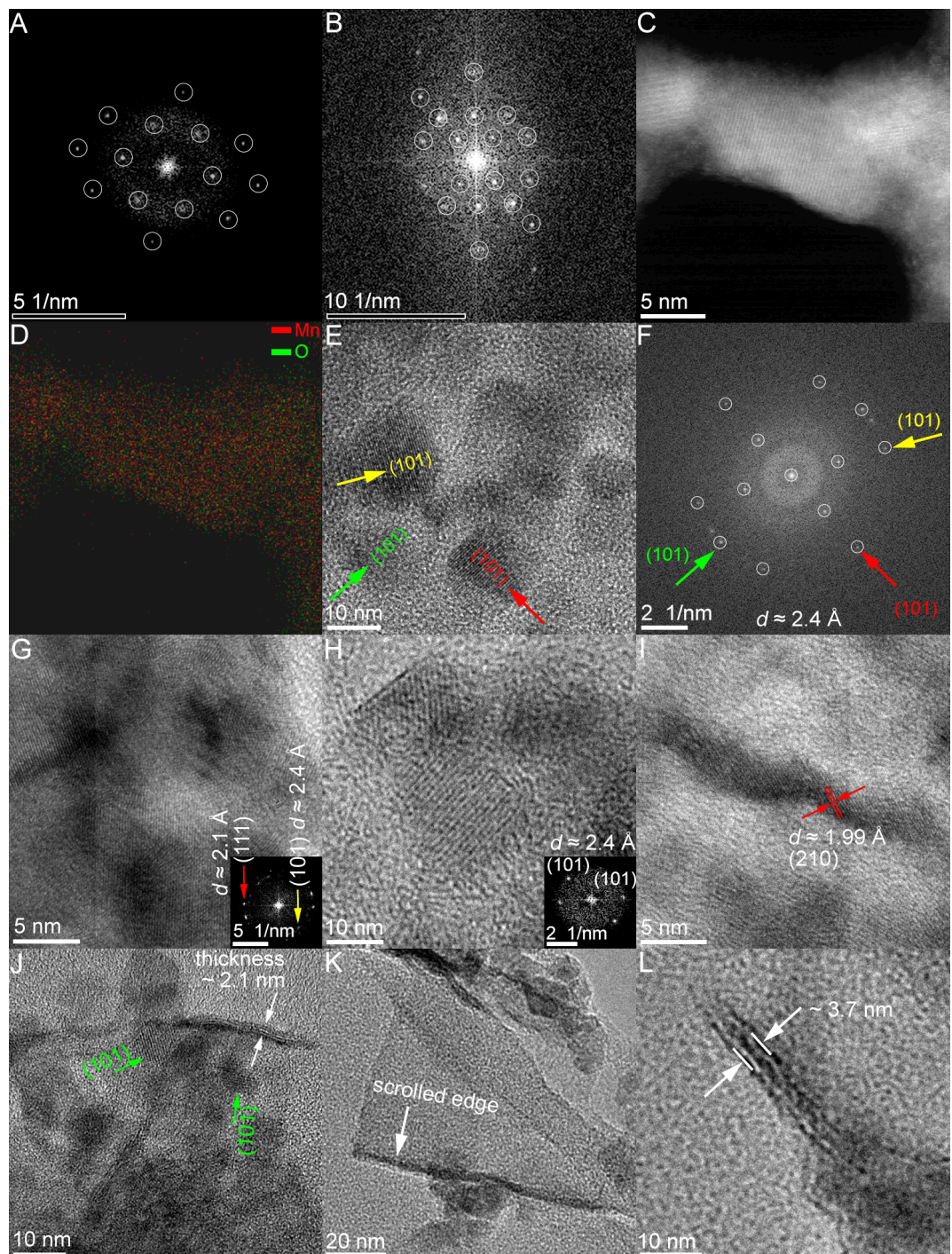

**Figure S4. TEM characterizations of $\beta$-MnO$_2$ NCs**

(A-B) FFT patterns corresponding with Figure 2E and F, respectively.

(C-D) TEM image (dark field) and corresponding element mapping.

(E-H) HRTEM images (bright field) and corresponding FFT patterns (FFT pattern of E is shown in F) suggested the dominant exposure of (101) and (111) (unzipping paralleling with the [001] zone axis).

(I-J) HRTEM images (bright field) of exfoliated MnO$_2$ edges.

(K-L) A MnO$_2$ sheet with scrolled edge (magnified in L. The sheet was moved during TEM characterization). The thickness of the scrolled and folded edge was around 3.7 nm, suggesting the ~1.85 nm thickness of the sheet.

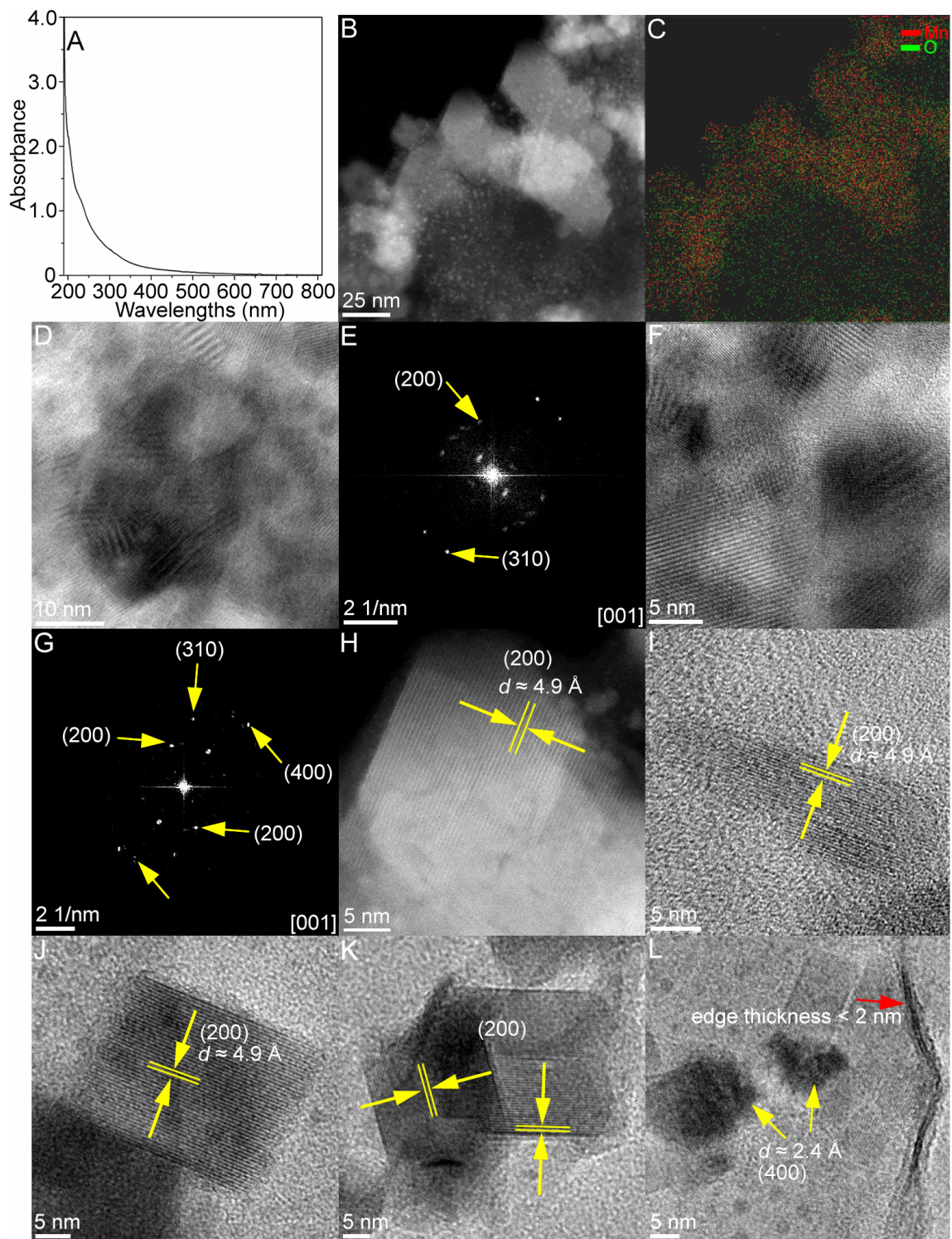

**Figure S5. UV/Vis spectrum and HRTEM images of *a*-MnO$_2$ NCs**

(A-B) Absorption spectrum.

(B-C) AC-TEM image and corresponding EDS mapping.

(D-G) (D,F) HRTEM images and (E,G) corresponding FFT patterns (e,g) suggested that the most exposed surface were the crystal facet which perpendicular to the [001] zone axis.

(H-L) HRTEM images show the (200) and (400) lattice. (H) is the dark field image.

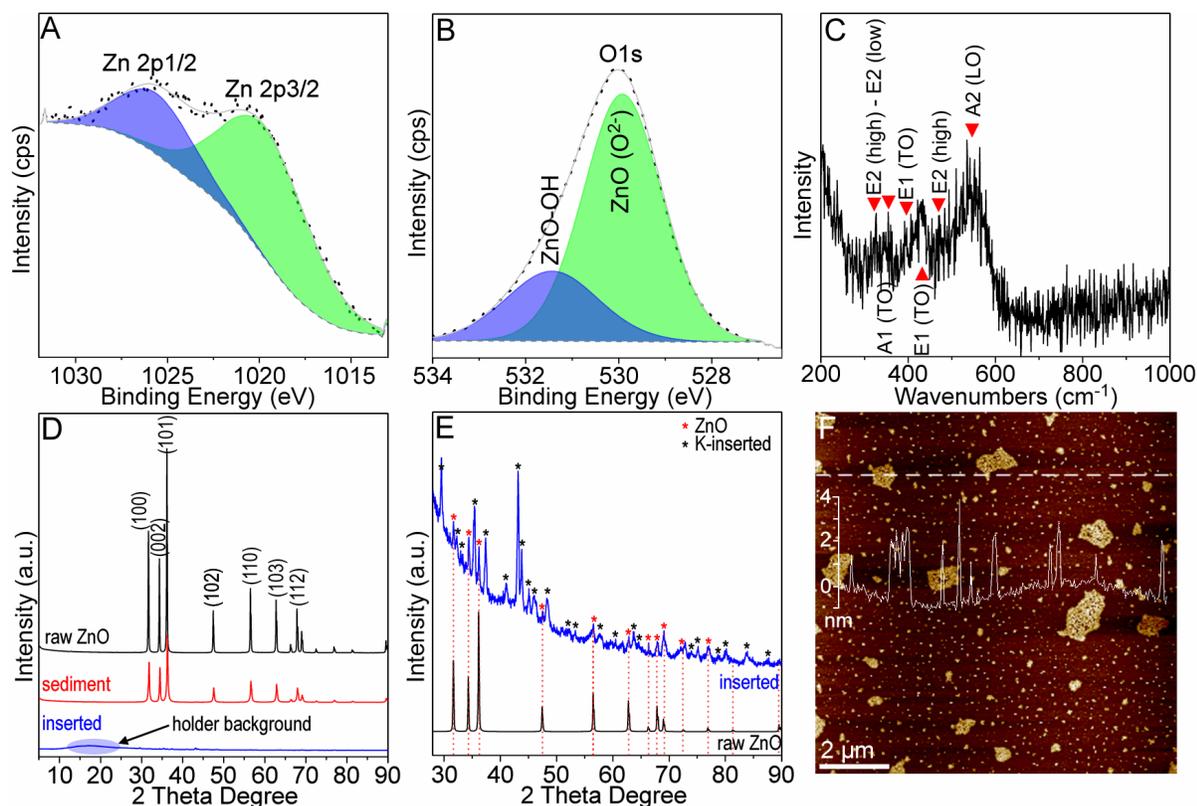

**Figure S6. Fabrication and Characterization of ZnO NCs**

(A-B) (A) Zn2p and (B) O1s XPS spectra of ZnO NCs.

(C) Raman spectrum

(D) XRD patterns of ZnO raw materials, sediment and K-ZnO with nearly same amount of the powder. Inset is the self-made XRD holder for the K-inserted sample.

(E) XRD pattern of ZnO raw materials and K-inserted ZnO. The relative XRD intensity was changed for better comparison.

(F) AFM topography of raw ZnO suspension, showing many 2D porous ZnO sheets.

**Additional information of Figure S6**: In Zn2p XPS spectrum (A), the Zn 2p1/2 and Zn 2p3/2 can be well recognized at 1026.2 and 1020.7 eV respectively.[S5] Except the ZnO, O1s XPS (B) suggest the surface –OH with a binding energy at around 531.5 eV.[S5] Raman spectrum (C) shows various modes at around 549, 470, 430-393 (430, 407, 393), 354 and 327 cm$^{-1}$, which can be assigned to the modes of A2 (LO), E2 (high), E1 (TO) (surface O vacancy and –OH groups), A1 (TO) and E2 (high) – E2 (low) of ZnO nanomaterials, respectively.[S6,S7] XRD patterns (D) suggest that the crystal structure of ZnO (ICCD card: 01-079-0207, hexagonal wurtzite) was highly disordered by the K-insertion (blue pattern), which was gently recovered after the extraction of K (red pattern). Close examination of the XRD patterns (E) suggest that the diffractions from pure potassium (ICCD card: 01-089-4080) was nearly invisible. Comparing with ZnO raw material, beside the weakened diffractions from ZnO structure, many new diffraction peaks appeared at lower angle positions, which all can be assigned to the K inserted ZnO materials (ICDD card: 01-076-0733 for $K_2ZnO_2$, 01-070-1276 and 01-070-0473 for $K_2Zn_6O_7$). AFM topography of raw ZnO suspension (F) gives many large porous sheets. The line profile suggests the thickness of around 2 nm of these sheets and dots, confirmed the thin 2D structure.

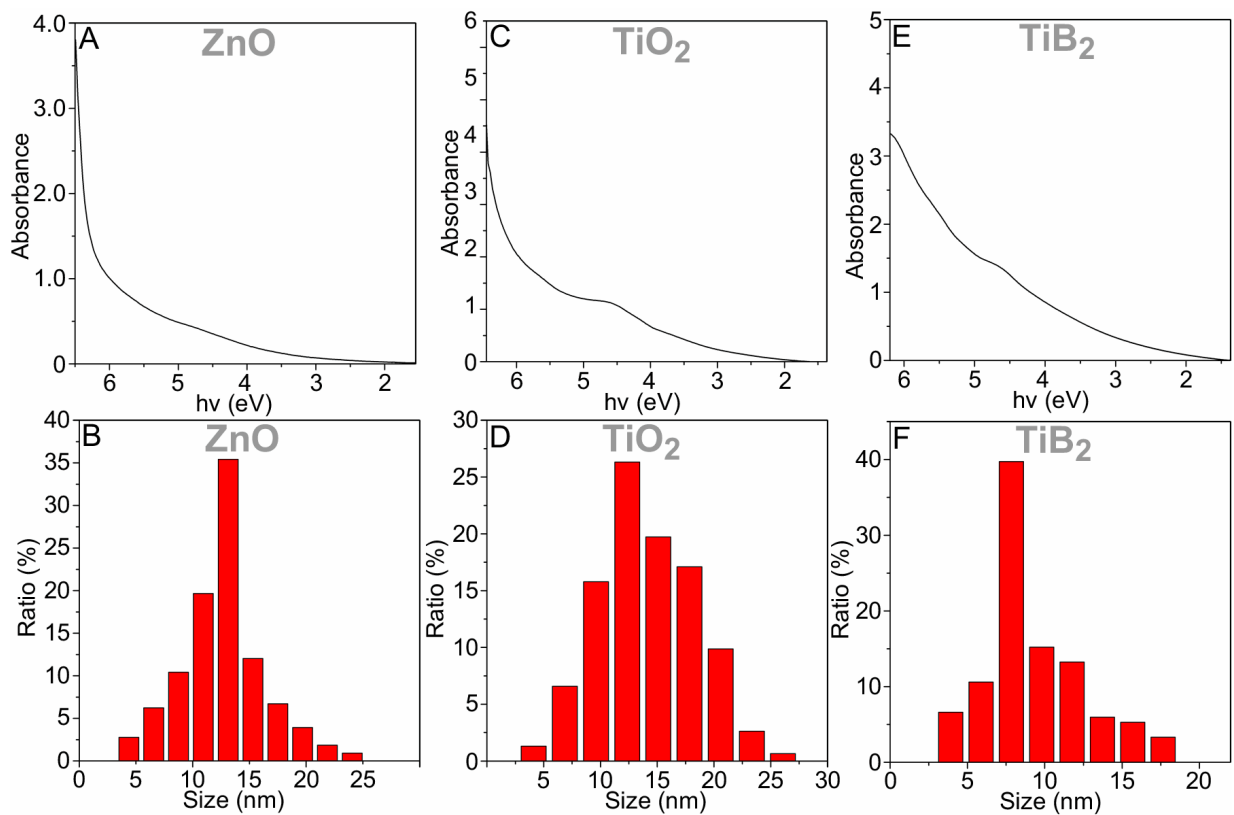

**Figure S7. UV/Vis spectrum (Tauc plot) and size distribution of $MnO_2$ NCs**

(A,C,E) Tauc plots of ZnO, $TiO_2$ and $TiB_2$.

(B,D,F) Size distribution of ZnO, $TiO_2$ and $TiB_2$ (from AFM images). The size from AFM image may be larger than the real size of NCs.

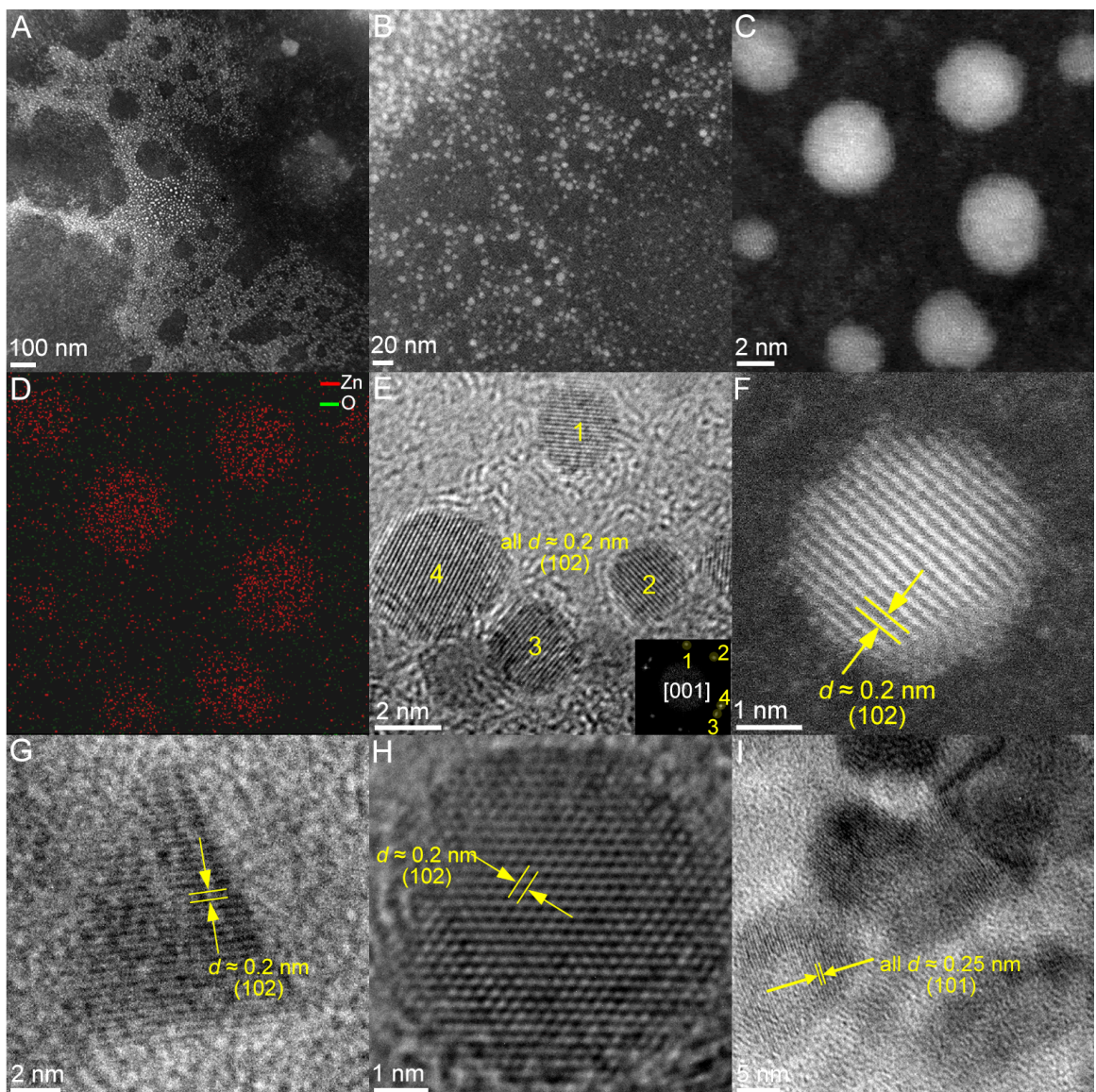

**Figure S8. TEM characterizations of ZnO NCs**

(A-D) AC-TEM images at dark field and corresponding elemental mapping (D is the elemental mapping of C, some O may from the carbon film used in the characterization). (E-I) HRTEM images (only F is the dark field AC-TEM image). Inset in E is the FFT pattern of different NCs.

**Additional information of Figure S8**: TEM images (A-B) suggest the small size of NCs (around or slightly smaller than 10 nm). Element mapping (C-D) of some NCs confirmed the Zn and O element. More HRTEM images (E-I) suggest the main surface exposure of NCs, which perpendiculars to the [010] zone axis of ZnO. The FFT pattern of some dot in one HRTEM image (E) shows only the diffraction from (102) plane. The lattices of (101) of wurtzite ZnO are also well showed in the HRTEM images (E-I). It should be noted that the main surface exposure was concluded also with more TEM images, but we only show some representing data here (same to the TEM images of other samples).

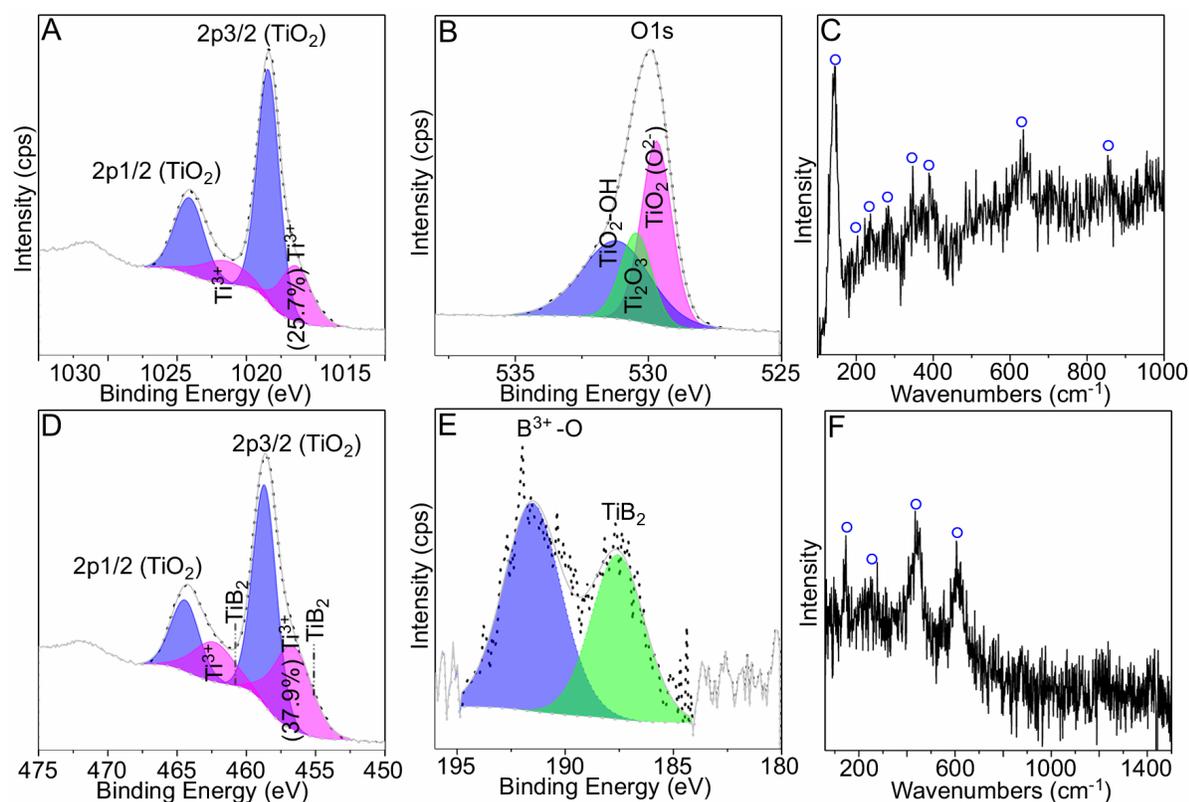

**Figure S9. XPS and Raman spectra of TiO₂ and TiB₂ NCs**

(A-B) (A) Ti2p and (B) O1s XPS spectra of TiO₂ NCs.
(C) Raman spectrum of TiO₂ NCs.
(D-E) (D) Ti2p and (E) B1s XPS spectra of TiB₂ NCs.
(F) Raman spectrum of TiB₂ NCs.

**Additional information of Figure S9**: In the XPS spectrum of TiO₂ NCs (A), the Ti2p can be fitted with $Ti^{4+}$ and $Ti^{3+}$ at (458.45 and 464.2 eV) and (456.5 and 461.7 eV), respectively. The ratio of $Ti^{3+}$ is around 25.7%. The O1s at 529.7, 530.5 and 530.3 eV can be assigned to $TiO_2$, $Ti_2O_3$ and non-lattice –OH groups (B). These values are the nearly the same to the reported data.[S8] Raman spectrum of TiO₂ NCs (C) shows peaks at around 144.8, 202.4, 235.8, 281.5, 346.7, 389.3, 633.8 cm⁻¹, which all can be assigned to the TiO₂ and defected phase.[S9] The Ti2p XPS spectrum of TiB₂ NCs can be fitted with $Ti^{4+}$ ($TiO_2$), $Ti^{3+}$ at (458.7 and 464.5 eV) and (456.7 and 462.5 eV), respectively (D). The ratio of $Ti^{3+}$ was around 37.9%. The Ti in TiB₂ was at the position with lower binding energy (D). In the B1s XPS spectrum of TiB₂ NCs (E), both the $B^{3+}$ and B states were identified at around 191.5 and 187.6 eV, respectively. These values are according well with the reported data.[S10] Raman spectrum of TiB₂ NCs gives peaks at around 142.9, 250, 437.9 and 605.7 cm⁻¹ (F). The latter three peaks can be assigned to the Raman modes from TiB₂.[S11,S12] The peak at 142.9 cm⁻¹ is close to that in TiO₂, which may induced by the oxidization phase in TiB₂.

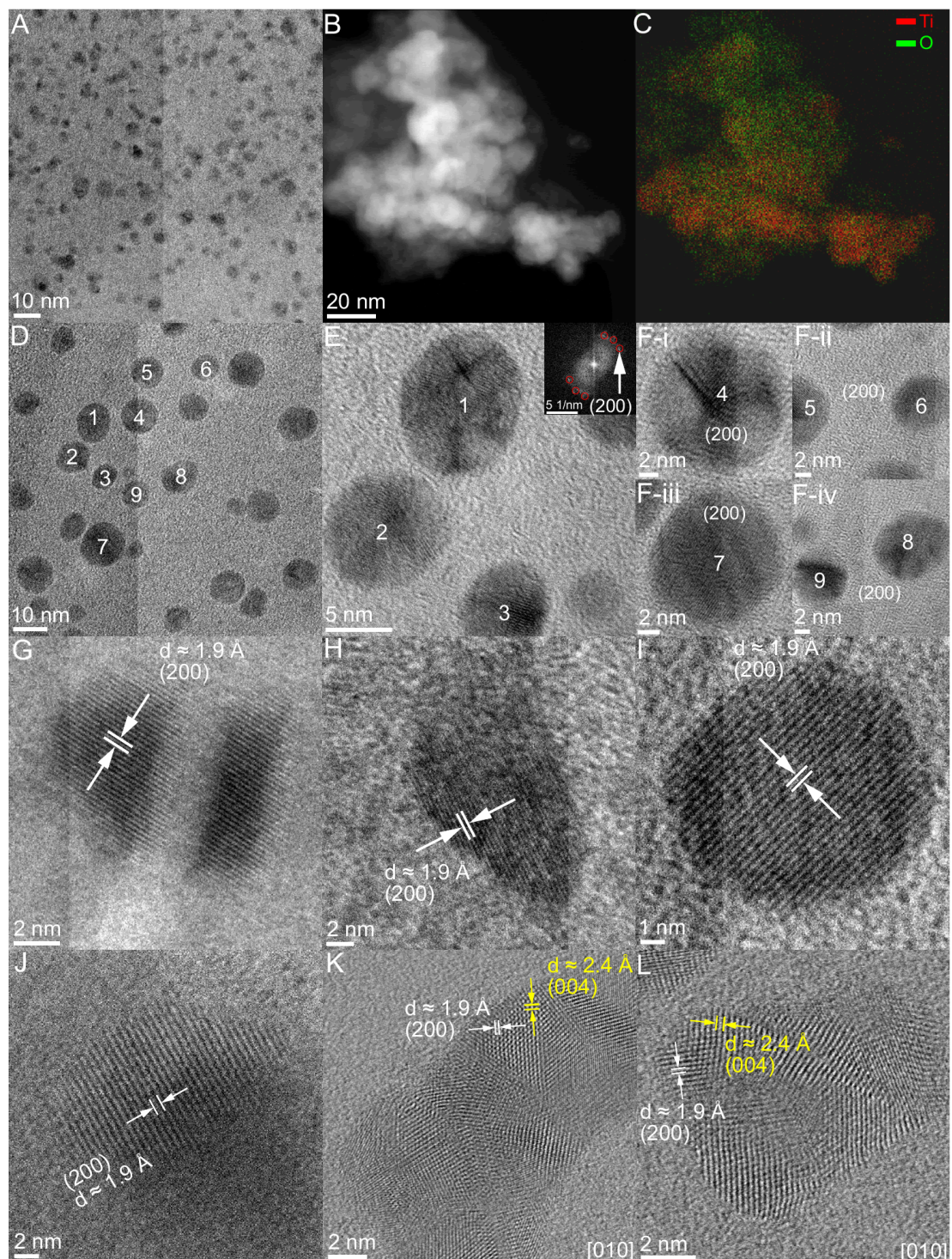

**Figure S10. TEM characterizations of TiO₂ NCs.**

(A) TEM image shows the size distribution.

(B-C) (B) TEM and (C) corresponding EDS mapping.

(D-L) TEM and HRTEM images. Inset in (E) is the FFT pattern, showing the only diffractions from (200).

**Additional information of Figure S10**: TEM images show the dominant (200) exposure of NCs. The labeled NCs in (D) are magnified in (E-F), which suggest that all these NCs have (200) exposures. Lattice fringes of (200) of TiO$_2$ were clearly identified (J-L, ICCD card: 03-065-5714), while the crystal structures in (K-L) were dislocated, giving slight different angles between lattices of (200) and (004).

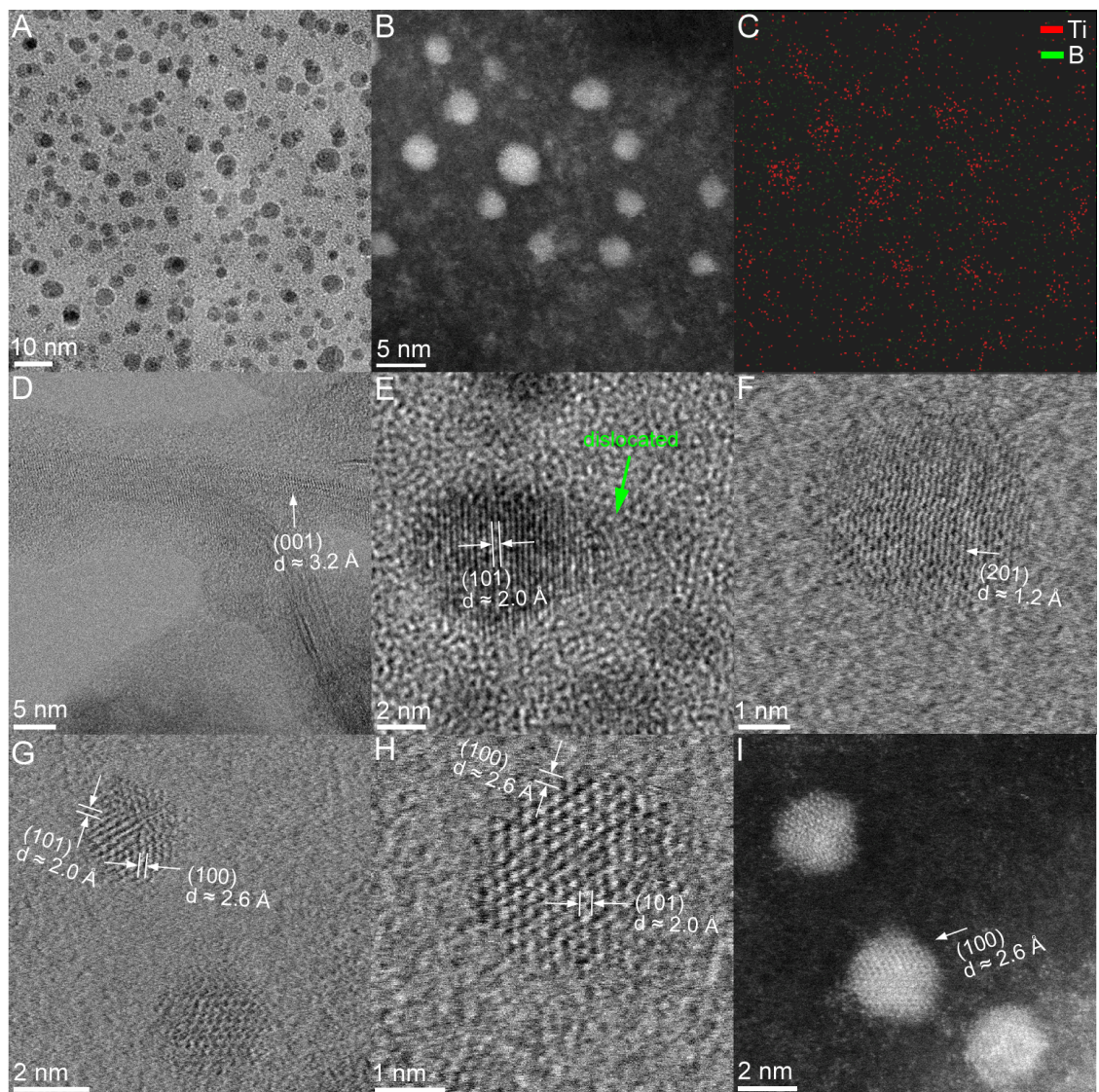

**Figure S11. TEM characterizations of TiB$_2$ NCs.**
(A) TEM image shows the size distribution.
(B-C) (B) TEM and (C) corresponding EDS mapping (the B signal was weak comparing with that of Ti).
(D) TEM image gives the (001) lattice of one nanosheet.
(E-I) TEM images show the crystal surface perpendicular to the [010] zone axis. (I) is the dark field image.

**Additional information of Figure S11**: From these TEM images, the surface exposure of NCs perpendicular to the [010] zone axis seems dominant. However, these NCs were mostly highly disordered, which may affected our analyses. We cannot conclude the main surface exposure of the TiB$_2$ NCs at current stage.

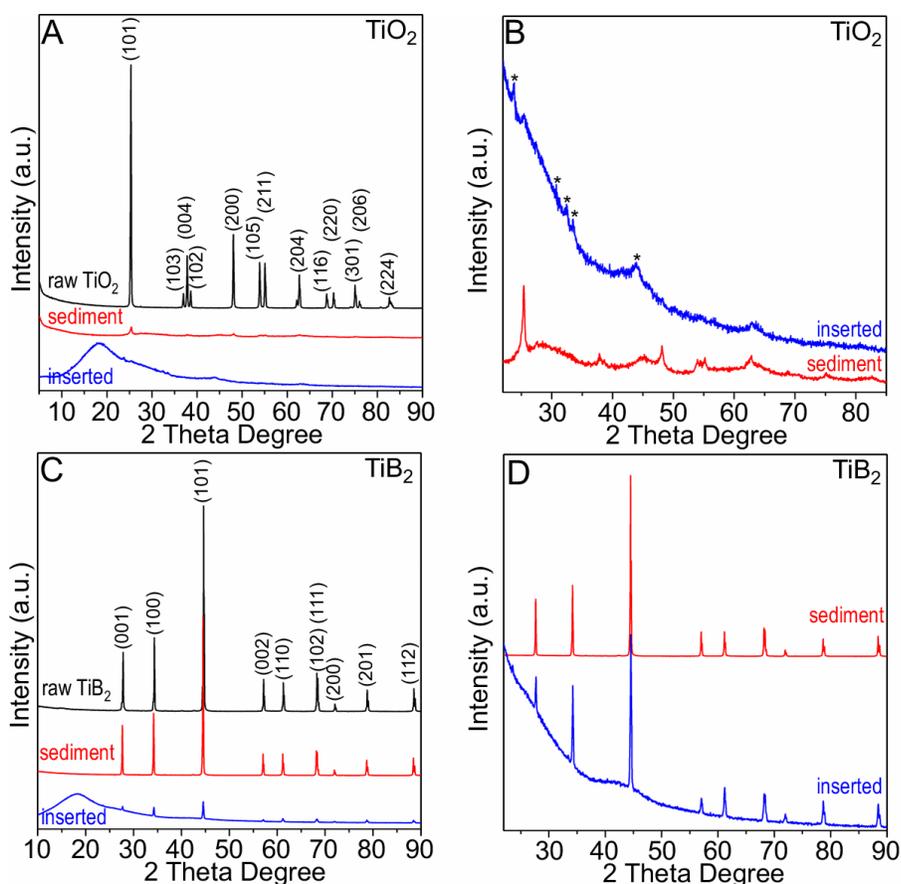

**Figure S12. XRD characterization of TiO₂ and TiB₂ materials**

(A) XRD patterns of TiO₂ raw materials, sediment and K-inserted TiO₂ powder with nearly same amount.

(B) XRD patterns of TiO₂ sediment and K-inserted TiO₂. The intensity of the pattern was changed for better comparison.

(C) XRD patterns of TiB₂ raw materials, sediment and K-inserted TiB₂ with nearly same amount.

(D) XRD patterns of K-TiB₂ and sediment. The intensity of the pattern was changed for better comparison.

**Additional information of Figure S12**: XRD patterns (A) suggest that the crystal structure of anatase TiO₂ (ICCD card: 03-065-5714) was highly disordered by K-insertion (blue pattern). The crystal structure of anatase TiO₂ was slightly recovered after the exfoliation reaction (red pattern). In the magnified XRD patterns (B), the peaks in TiO₂ sediment were weak, but the anatase phase was identified (ICCD card: 03-065-5714). Besides the very weak (close to the instrument background) XRD diffractions, other new peaks have also been recognized (asterisk), which can be associated to the K-inserted structure (*e.g.* the ICCD cards of 01-081-2039, 00-001-1016 and 00-013-0447). XRD patterns (C) suggest that the crystal structure of TiB₂ (ICCD card: 00-035-0741) was highly disordered in K-TiB₂, which was generally recovered in sediment. In the magnified XRD patterns (D), unlike other materials in the work, the presence of relatively strong XRD diffractions of TiB₂ in K-TiB₂ suggested that the K has been inserted into part of the structure, *i.e.* part of the structure has been fully disordered and others were still remained as highly crystallized.

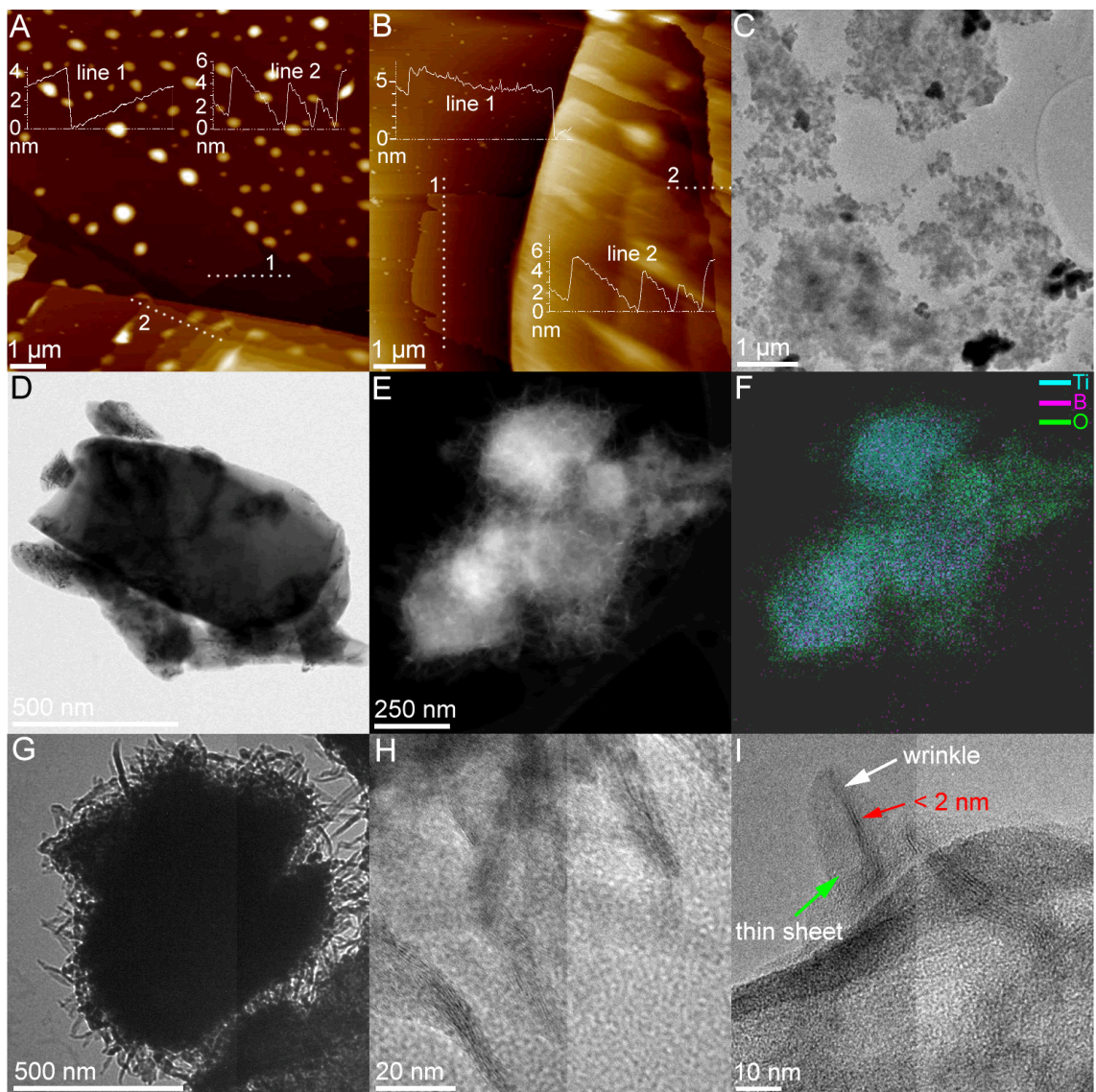

**Figure S13. AFM and TEM characterizations of TiO$_2$ and TiB$_2$ materials**

(A-B) AFM images and corresponding 2D line profile of the residual TiO$_2$ particles after one time fabrication, showing some ultrathin edges and crystal steps.

(C) TEM image shows the cracked TiO$_2$ structures.

(D) TEM image of the raw TiB$_2$ particle.

(E-F) One AC-TEM image and corresponding element mapping of residual TiB$_2$ particles after 5 times fabrication.

(G-I) TEM images of the surface of residual TiB$_2$ particle.

**Additional information of Figure S13**: Comparing with the raw TiB$_2$ particle (D), the surface of residual TiB$_2$ was highly exfoliated and disordered, giving many crystal whiskers which have not been exfoliated into the NCs suspension (E,G). The whiskers were very thin, giving some wrinkles on the sheet (H). One whisker shows the thickness of < 2 nm at the wrinkle position, which suggested that the whiskers were very thin (I).

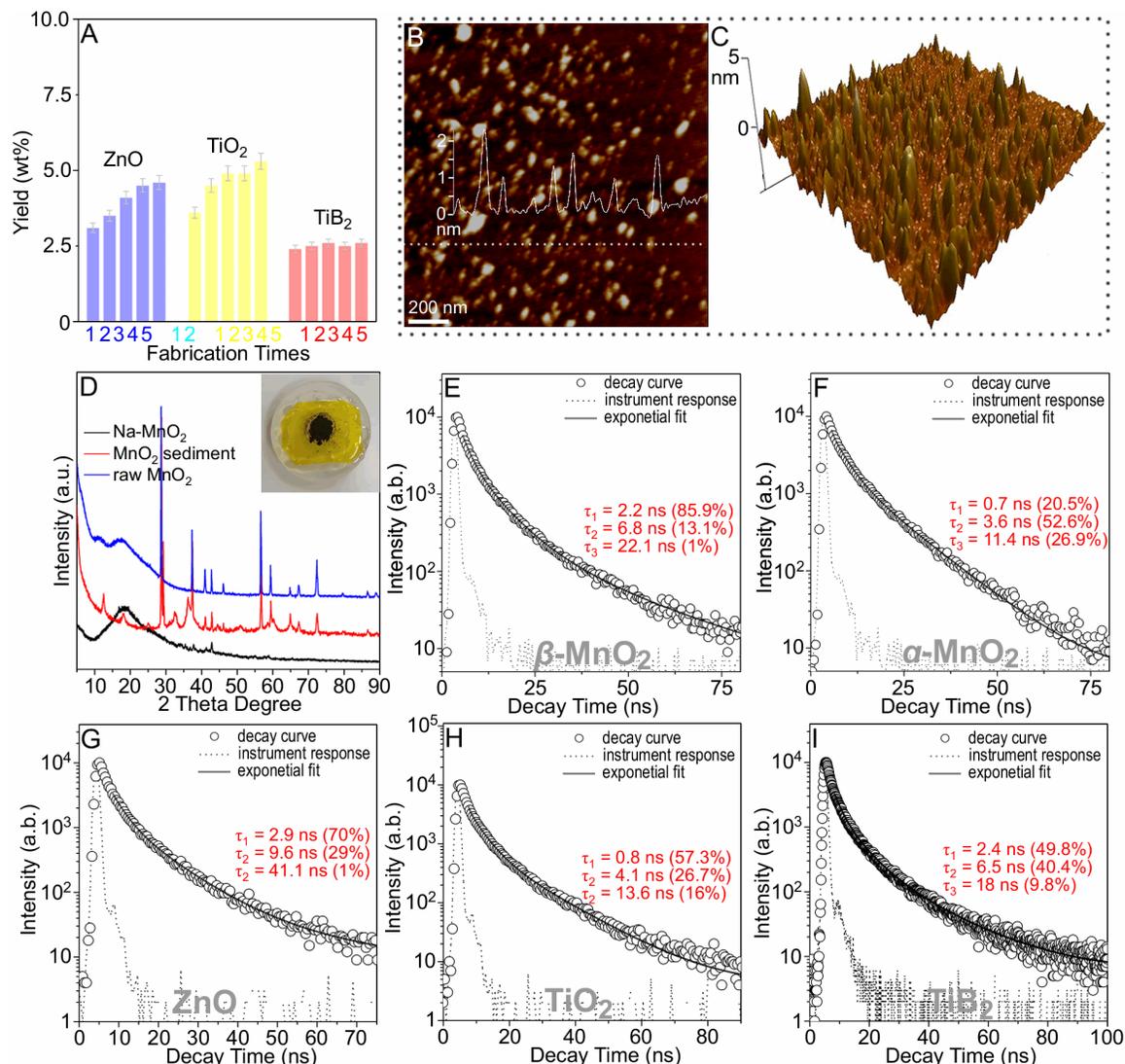

**Figure S14. Additional fabrications and optical properties of NCs**

(A) A bar chart of the yield of different NCs against the fabrication time. The error bar was treated as ±2.5 wt% generated in the purification process.

(B,C) AFM images (2D profile, B, and 3D profile, C) of $MnO_2$ NCs fabricated with Na intercalant.

(D) XRD patterns of the $MnO_2$ raw material, Na-$MnO_2$ and corresponding sediment. Inset is the Na-$MnO_2$ sample.

(E) TRPL of $a$-$MnO_2$ NCs.

(F) TRPL of $\beta$-$MnO_2$ NCs.

(G) TRPL of ZnO NCs.

(H) TRPL of $TiO_2$ NCs.

(I) TRPL of $TiB_2$ NCs.

**Additional information of Figure S14**: In figure D, all the diffraction peaks of the $MnO_2$ sediment can be well assigned to $a$-$MnO_2$ (ICDD card: 01-072-1982), which has many similar peaks with $\beta$-$MnO_2$ (ICDD card: 01-081-2261) above 20°. In the TRPL measurement, laser excitation wavelength of 340 nm was used, with light emission collected was at different wavelengths, corresponding to their respective emission peaks, as collected from steady state PL.

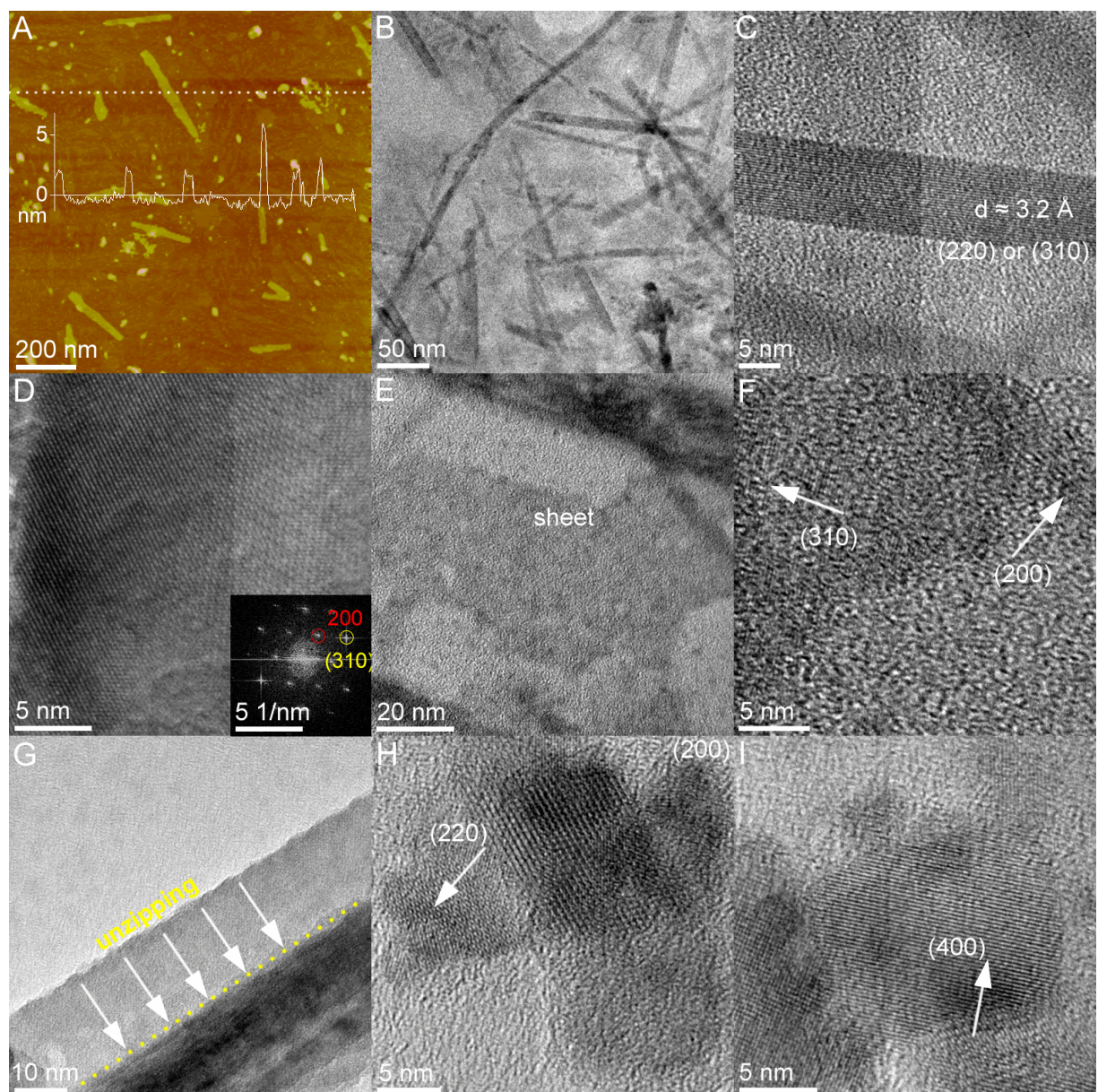

**Figure S15. AFM and TEM characterizations of MnO₂ fabricated with Na intercalant.**
(A) AFM image.
(B-I) TEM images

**Additional information of Figure S15**: AFM image (A) shows many ultrathin nanoribbons and nanoparticles, which is further revealed by the TEM image (B). The nanoribbons are highly crystallized. Analysis of the TEM image (together with FFT pattern) suggested that these nanoribbons were mainly a-MnO₂ (ICCD card: 01-072-1982, C,D). Occasionally, some sheets were also found in the samples (E). The sheet in E is also crystallized (F). G shows a partially unzipped ribbon. These samples also contained many small nanocrystals which are also a-MnO₂ (H,I).

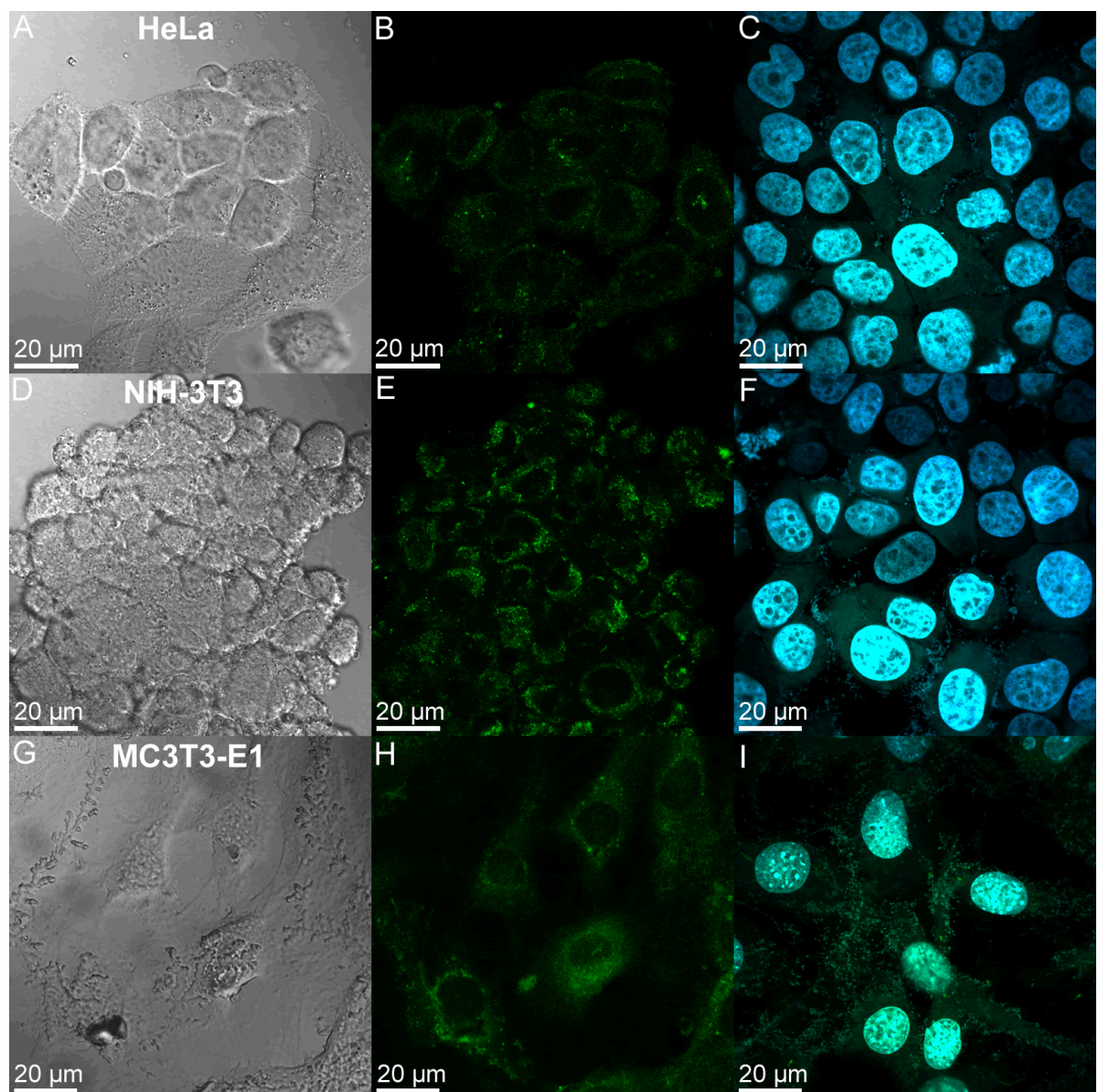

**Figure S16. Bio-sensing of TiB$_2$ NCs with different cells**

(A-B) Confocal microscopy images of HeLa cells stained with TiB$_2$ NCs, (A) without and (B) with single light excitation.

(C) Confocal microscopy image of HeLa cells stained with both DAPI and TiB$_2$ NCs (merged from two images with different light excitations).

(D-E) Confocal microscopy images of NIH-3T3 cells stained with TiB$_2$ NCs, (D) without and (E) with single light excitation.

(F) Confocal microscopy image of NIH-3T3 cells stained with both DAPI and TiB$_2$ NCs (merged from two images with different light excitations).

(G-H) Confocal microscopy images of MC3T3-E1 cells stained with TiB$_2$ NCs, (G) without and (H) with single light excitation.

(F) Confocal microscopy image of MC3T3-E1 cells stained with both DAPI and TiB$_2$ NCs (merged from two images with different light excitations).

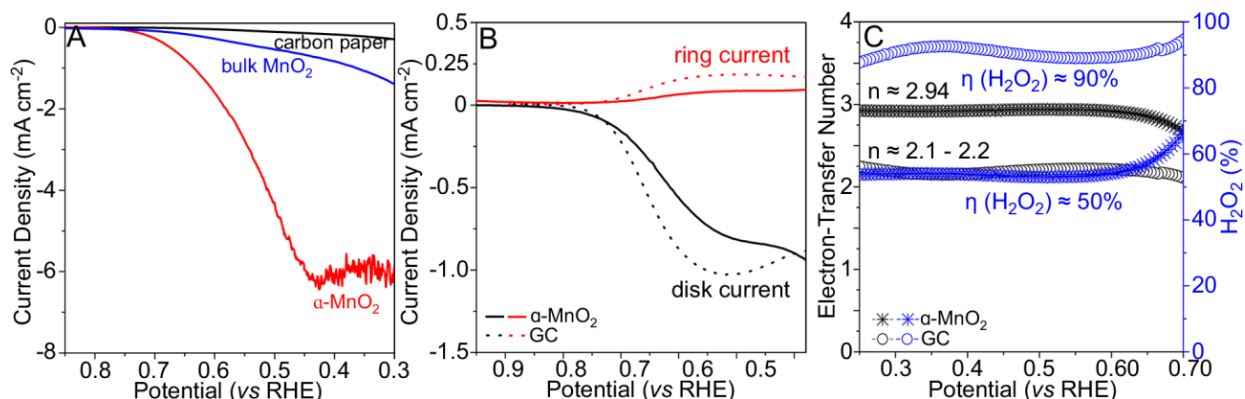

**Figure S17. Electrocatalysis measurement of MnO₂ materials**
(A) ORR LSV of carbon paper, bulk MnO₂ raw materials and *a*-MnO₂ NCs.
(B) LSV of the ring and disk current of the *a*-MnO₂ NCs.
(C) Plots of the electron transfer number and ratio of H₂O₂ production against the potential. Data was calculated from (B).

**Additional information of Figure S17**: The stacking of NCs significantly reduced the current density. Therefore, dispersed NCs were loaded on the carbon cloth to reach the highly exposed surface area and avoid the stacking of NCs as much as possible (A). For the measurement of electrode transfer number and the yield of the H₂O₂ with RRDE technique (B,C), NCs have to be directly mounted on the glassy carbon (GC) disk. We applied thick NCs to fully cover the GC surface. Although the severe stacking of NCs was unavoidable, this treatment allowed the correct measurement of the electron transfer number and the yield of H₂O₂ at this stage. From this measurement, the H₂O₂ yield of the GC electrode was high to be around 90%. By contrast, this value was decreased to 50% in the *a*-MnO₂ NCs.